\documentclass[prd,preprint,tightenlines,floatfix,showpacs,preprintnumbers,nofootinbib,eqsecnum]{revtex4}

 \usepackage[dvips,final]{graphicx}
  \usepackage{amssymb}
   \usepackage{amsmath}
    \usepackage{amsfonts}
     \usepackage{epsfig}
      \usepackage{bm}

\usepackage[section]{placeins}

\usepackage{multirow}
\usepackage{ctable}
\usepackage{booktabs}
\usepackage{array}
\usepackage{tabularx}
\usepackage{xcolor}
\usepackage{pstricks}

\begin{document}

\vfill
\title{Double-parton scattering contribution to production of jet pairs with large rapidity separation at the LHC}

\author{Rafa{\l} Maciu{\l}a}
\email{rafal.maciula@ifj.edu.pl} \affiliation{Institute of Nuclear Physics PAN, PL-31-342 Cracow, Poland}

\author{Antoni Szczurek}
\email{antoni.szczurek@ifj.edu.pl} \affiliation{Institute of Nuclear Physics PAN, PL-31-342 Cracow, Poland and\\
University of Rzesz\'ow, PL-35-959 Rzesz\'ow, Poland}

\date{\today}

\begin{abstract}

We discuss production of four-jet final state $pp \rightarrow j j j j X$
in proton-proton collisions at the LHC through the mechanism of
double-parton scattering (DPS) in the context of jets with large
rapidity separation.
The DPS contributions are calculated within the so-called factorized
Ansatz and each step of DPS is calculated in the LO collinear approximation. 
The LO pQCD calculations are shown to give a reasonably good
descritption of recent CMS and ATLAS data on inclusive jet
production and therefore this formalism can be used to reliably estimate
the DPS effects.
Relative contribution of DPS is growing at large rapidity distance between jets. 
This is consistent with our experience from previous studies of
double-parton scattering effects in the case of open and hidden charm production.
The calculated differential cross sections as a function of rapidity
distance between the most remote in rapidity jets are compared with 
recent results of LL and NLL BFKL calculations for Mueller-Navelet (MN) 
jet production at $\sqrt{s} = 7$ TeV. 
The DPS contribution to widely rapidity separated jet production 
is carefully studied for the present energy $\sqrt{s}$ = 7 TeV and also
at the nominal LHC energy $\sqrt{s}$ = 14 TeV and in different ranges 
of jet transverse momenta.
The differential cross section as a function of dijet transverse momenta
as well as two-dimensional ($p_{T}(y_{min})\times p_{T}(y_{max})$)-plane
correlations for DPS mechanism are also presented. Some ideas how the
DPS effects could be studied in the case of double dijet production 
are suggested.
\end{abstract}


\maketitle

\section{Introduction}

\begin{figure}[!h]
\begin{minipage}{0.35\textwidth}
 \centerline{\includegraphics[width=1.0\textwidth]{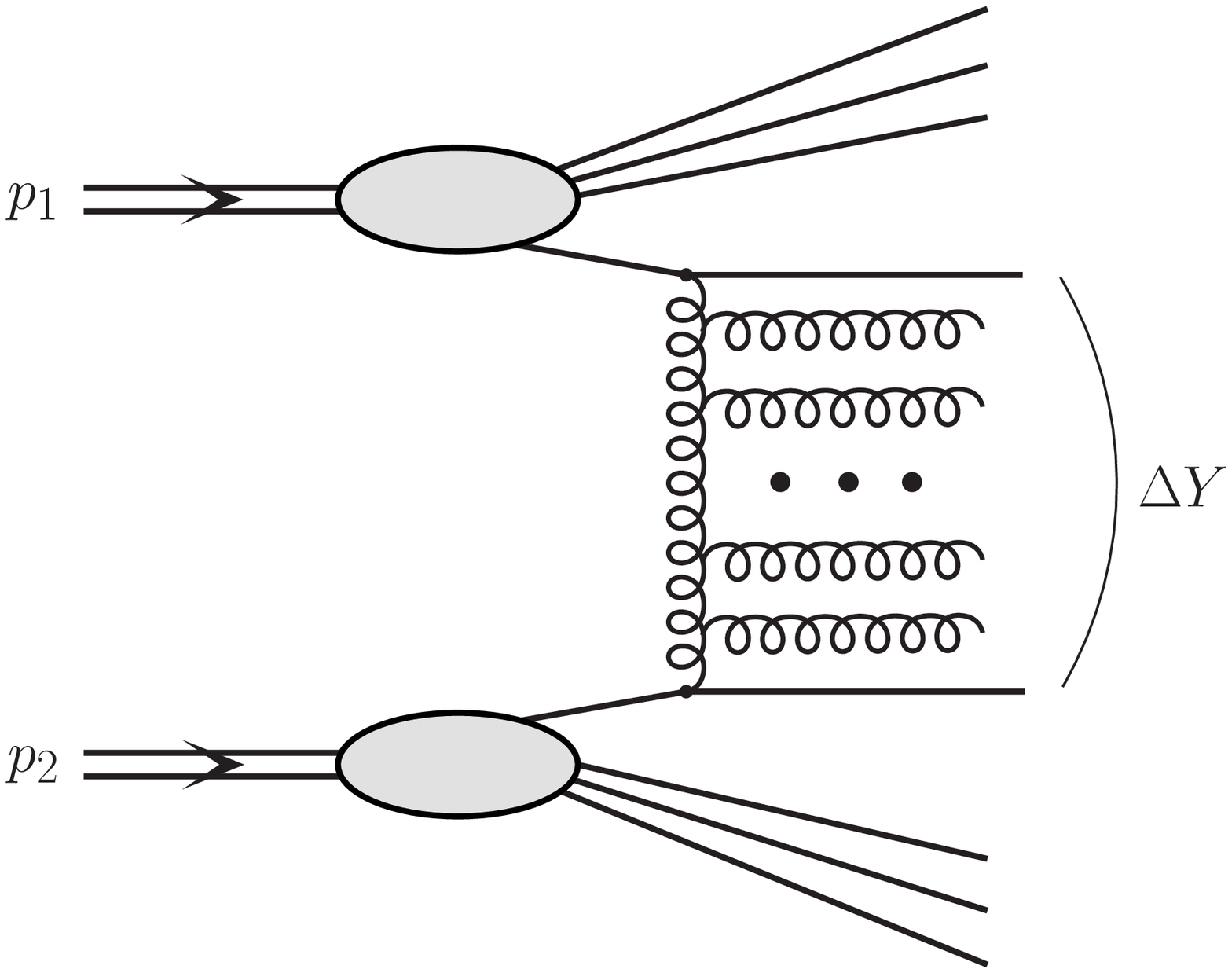}}
\end{minipage}
\hspace{0.5cm}
\begin{minipage}{0.38\textwidth}
 \centerline{\includegraphics[width=1.0\textwidth]{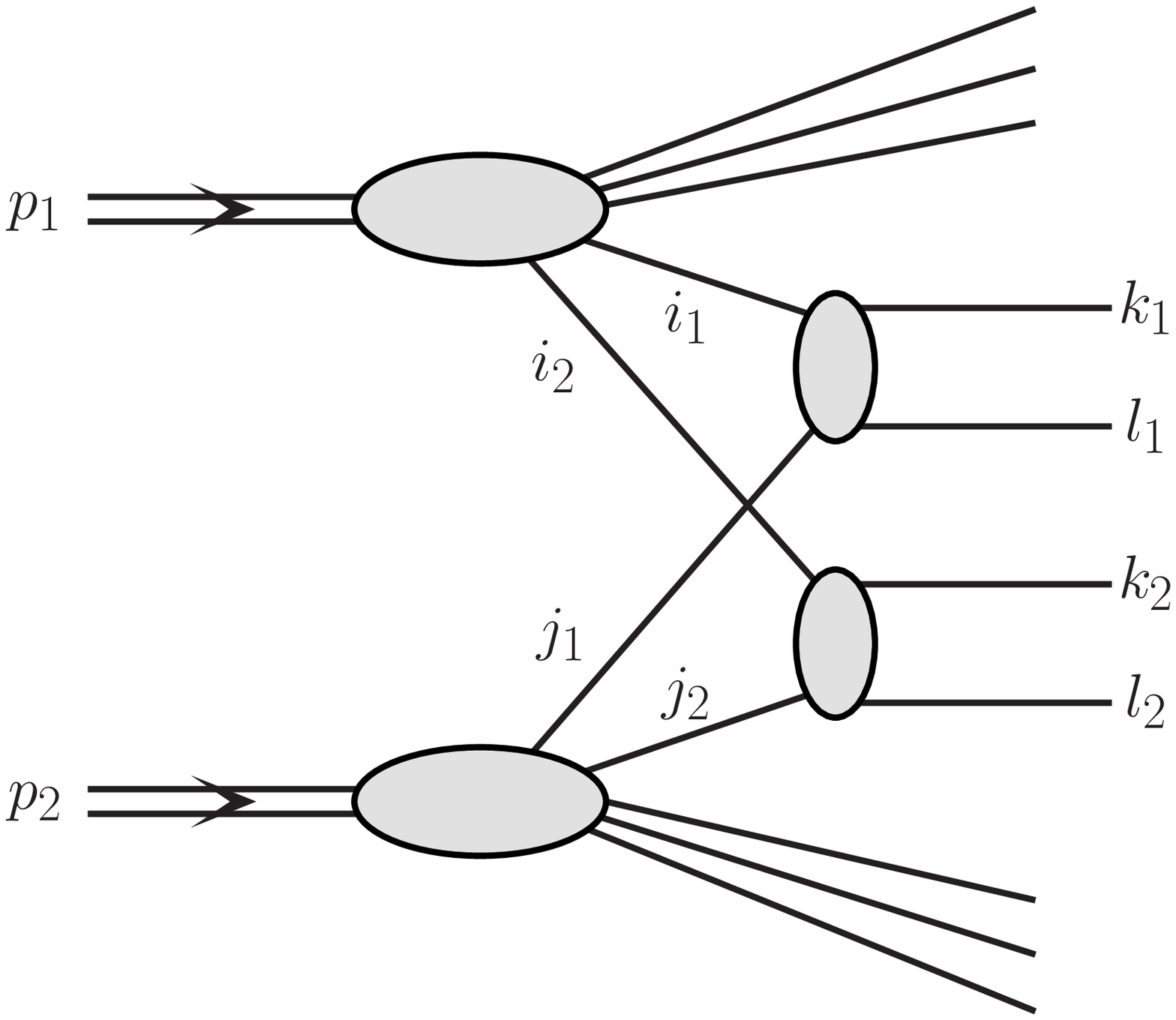}}
\end{minipage}
   \caption{
\small A diagramatic representation of the Mueller-Navelet jet
production (left diagram) and of the double paron scattering mechanism
(right diagram).
}
 \label{fig:diagrams}
\end{figure}

It is reasonable to expect that large-rapidity-distance jets are more
decorrellated in azimuth than jets placed close in rapidity. 
About 25 years ago Mueller and Navelet predicted strong decorrelation 
in relative azimuthal angle \cite{Mueller:1986ey} of such jets due to exchange 
of the BFKL ladder between quarks (partons). 
The generic picture is presented in diagram (a) of Fig.~\ref{fig:diagrams}.
In a bit simplified picture quarks/antiquarks/gluons are emitted forward
and backward, whereas gluons emitted along the ladder populate 
rapidity regions in between.
Due to diffusion along the exchange ladder the correlation
between the most forward and the most backward jets is small.
This was a simple picture obtained within leading-logarithmic 
BFKL formalism
\cite{Mueller:1986ey,DelDuca:1993mn,Stirling:1994he,DelDuca:1994ng,Kim96,Andersen2001}.
In Ref.~\cite{Kwiecinski:2001nh} so-called consistency constrain was imposed
in addition.
Recent higher-order BFKL calculation slightly modified
this simple picture 
\cite{Bartels-MNjets,Vera:2007kn,Marquet:2007xx,Colferai:2010wu,Caporale:2011cc,Ivanov:2012ms,
Caporale:2012ih,Ducloue:2013hia,Ducloue:2013bva,DelDuca2014} leading to smaller azimuthal decorrelation
in rapidity. Recently the NLL corrections were calculated
both to the Green's function and to the jet vertices.
The effect of the NLL correction is large and leads to significant
lowering of the cross section.
So far only averaged values of $<\!\!cos(n \phi_{jj})\!\!>$ 
over available phase space or even their ratios 
were studied experimentally \cite{CMS_MN1}. More detailed studies are necessary
to verify this type of calculations. In particular, the approach
should reproduce dependence on the rapidity distance between
the jets emitted in opposite hemispheres and more detailed 
associated dependences on transverse momenta of the jets.
Large-rapidity-distance jets can be produced only at high energies
where the rapidity span is large due to kinematics.
A first experimental trial of search for the MN jets was made by 
the D0 collaboration \cite{Abachi96}. In their study rapidity distance between jets was
limited to 5.5 units only.
Nonetheless they have observed a broadening of the $\phi_{jj}$
distribution with growing rapidity distance between jets.
However, theoretical interpretation of the broadening is not clear.
The dijet azimuthal correlations were also studied in collinear
next-to-leading order approximation \cite{Aurenche:2008dn}.
The LHC opens new possibility to study the decorrelation effect 
quantitatively.
First experimental data measured at $\sqrt{s}$ = 7 TeV are expected 
soon \cite{CMS_private}.

On the other hand recent studies of multiparton interactions have shown
that they may easily produce objects which are emitted 
far in rapidity.
Good example is production of $c \bar c c \bar c$ 
\cite{Luszczak:2011zp,Maciula:2013kd,Hameren2014}
or inclusive production of two $J/\psi$ mesons
\cite{Kulesza-Stirling,Baranov:2012re}. 
Here we wish to concentrate on four-jet double-parton scattering (DPS)
production with large distances between jets (see diagram (b)
in Fig.~\ref{fig:diagrams}).
Several suggestions how to separate four-jet DPS contribution from SPS
contribution at midrapidities were discussed in Ref.~\cite{Berger:2009cm}.

In the present first exploratory study we shall make first estimate of 
the DPS effects for jets with large rapidity separation 
within leading-order collinear approximation. Already this approximation
will allow us to nicely illustrate the generic situation. We shall focus 
on distribution in rapidity distance of the most-rapidity-distant jets.
The DPS result will be compared to the distribution in rapidity distance
for standard 2 $\to$ 2 single parton scattering (SPS) pQCD dijet 
calculation. We shall identify the dominant partonic subprocesses 
important to understand the situation in the small but interesting corner 
of the phase space.
The calculation of distributions in rapidity distance will be
supplemented by the analysis of correlations in the two-dimensional
space of the transverse momenta of the two widely separated jets or 
by calculation of distributions in transverse momentum imbalance of 
the jets or correlations in azimuthal angle between them.

\section{Basic formalism}

In the present calculation all partonic cross sections ($i j \to k l$) are calculated in
leading-order only.
The cross section for dijet production can be written then as:
\begin{equation}
\frac{d \sigma(i j \to k l)}{d y_1 d y_2 d^2p_t} = \frac{1}{16 \pi^2 {\hat s}^2}
\sum_{i,j} x_1 f_i(x_1,\mu^2) \; x_2 f_j(x_2,\mu^2) \;
\overline{|\mathcal{M}_{i j \to k l}|^2} \;,
\label{LO_SPS}
\end{equation}
where $y_1$, $y_2$ are rapidities of the two jets ($k$ and $l$) and $p_t$ is
transverse momentum of one of them (they are identical). The parton distributions are evaluated
at $x_{1} = \frac{p_{t}}{\sqrt{s}} (\exp{(y_1)}+\exp{(y_2)}), x_{2} = \frac{p_{t}}{\sqrt{s}} (\exp{(-y_1)}+\exp{(-y_2)})$ and $\mu^{2} = p_{t}^{2}$ is used as factorization and renormalization scale.

In our calculations we include all leading-order $i j \to k l$ partonic 
subprocesses (see e.g. \cite{Ellis-Stirling-Webber,Barger-Phillips}).
The $K$-factor for dijet production is rather small, of the order of 
$1.1 - 1.3$ (see e.g. \cite{K-factor1,K-factor2}), 
but can be easily incorporated in our calculations. Below we shall show that
already the leading-order approach gives results in sufficiently reasonable 
agreement with recent ATLAS \cite{ATLASjets} and CMS \cite{CMSjets} data.

This simplified leading-order approach can, however be, used conveniently
in our first estimate of DPS differential cross sections for jets 
widely separated in rapidity. 
In analogy to the production of $c \bar c c \bar c$ (see e.~g.~\cite{Luszczak:2011zp}) one can write:
\begin{equation}
\frac{d \sigma^{DPS}(p p \to \textrm{4jets} \; X)}{d y_1 d y_2 d^2p_{1t} d y_3 d y_4 d^2p_{2t}} = \sum\nolimits_{\substack{i_1,j_1,k_1,l_1\\ i_2,j_2,k_2,l_2}} \; \frac{\mathcal{C}}{\sigma_{eff}} \;
\frac{d \sigma(i_1 j_1 \to k_1 l_1)}{d y_1 d y_2 d^2p_{1t}} \; \frac{d \sigma(i_2 j_2 \to k_2 l_2)}{d y_3 d y_4 d^2p_{2t}}\;, 
\label{DPS}
\end{equation}
where
$\mathcal{C} = \left\{ \begin{array}{ll}
\frac{1}{2}\;\; & \textrm{if} \;\;i_1 j_1 = i_2 j_2 \wedge k_1 l_1 = k_2 l_2\\
1\;\;           & \textrm{if} \;\;i_1 j_1 \neq i_2 j_2 \vee k_1 l_1 \neq k_2 l_2
\end{array} \right\} $ and partons $i,j,k,l = g, u, d, s, \bar u, \bar d, \bar s$. 
The combinatorial factors include identity of the two subprocesses.
Each step of the DPS is calculated in the leading-order approach (see
Eq.(\ref{LO_SPS})).
The quantity $\sigma_{eff}$ has dimension of cross section and has
a simple interpretation in the impact parameter representation
\cite{Gustaffson2011}.
Above $y_1$, $y_2$ and $y_3$, $y_4$ are rapidities of partons in
first and second partonic subprocess, respectively.
The $p_{1t}$ and $p_{2t}$ are respective transverse momenta.

Experimental data from the Tevatron \cite{Tevatron} and the LHC 
\cite{Aaij:2011yc,Aaij:2012dz,Aad:2013bjm} 
provide an estimate of $\sigma_{eff}$ in the denominator of formula 
(\ref{DPS}). As in our recent paper \cite{Hameren2014} we take 
$\sigma_{eff}$ = 15 mb.
A detailed analysis of $\sigma_{eff}$ based on various experimental data
can be found in Refs.~\cite{Seymour:2013qka,Bahr:2013gkj}.

\section{Numerical Results}
\label{sec:results}

\begin{figure}[!h]
\begin{minipage}{0.47\textwidth}
 \centerline{\includegraphics[width=1.0\textwidth]{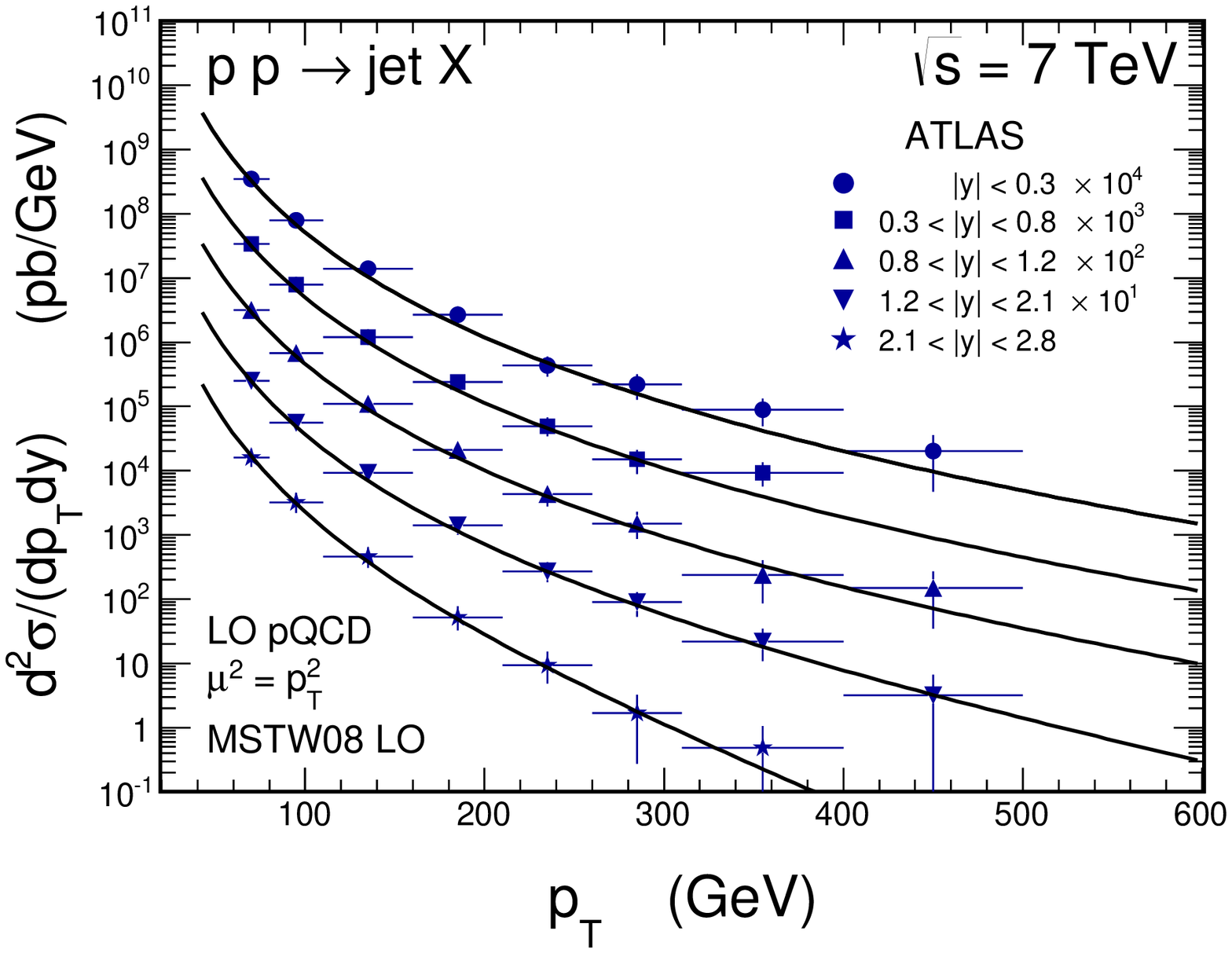}}
\end{minipage}
\hspace{0.5cm}
\begin{minipage}{0.47\textwidth}
 \centerline{\includegraphics[width=1.0\textwidth]{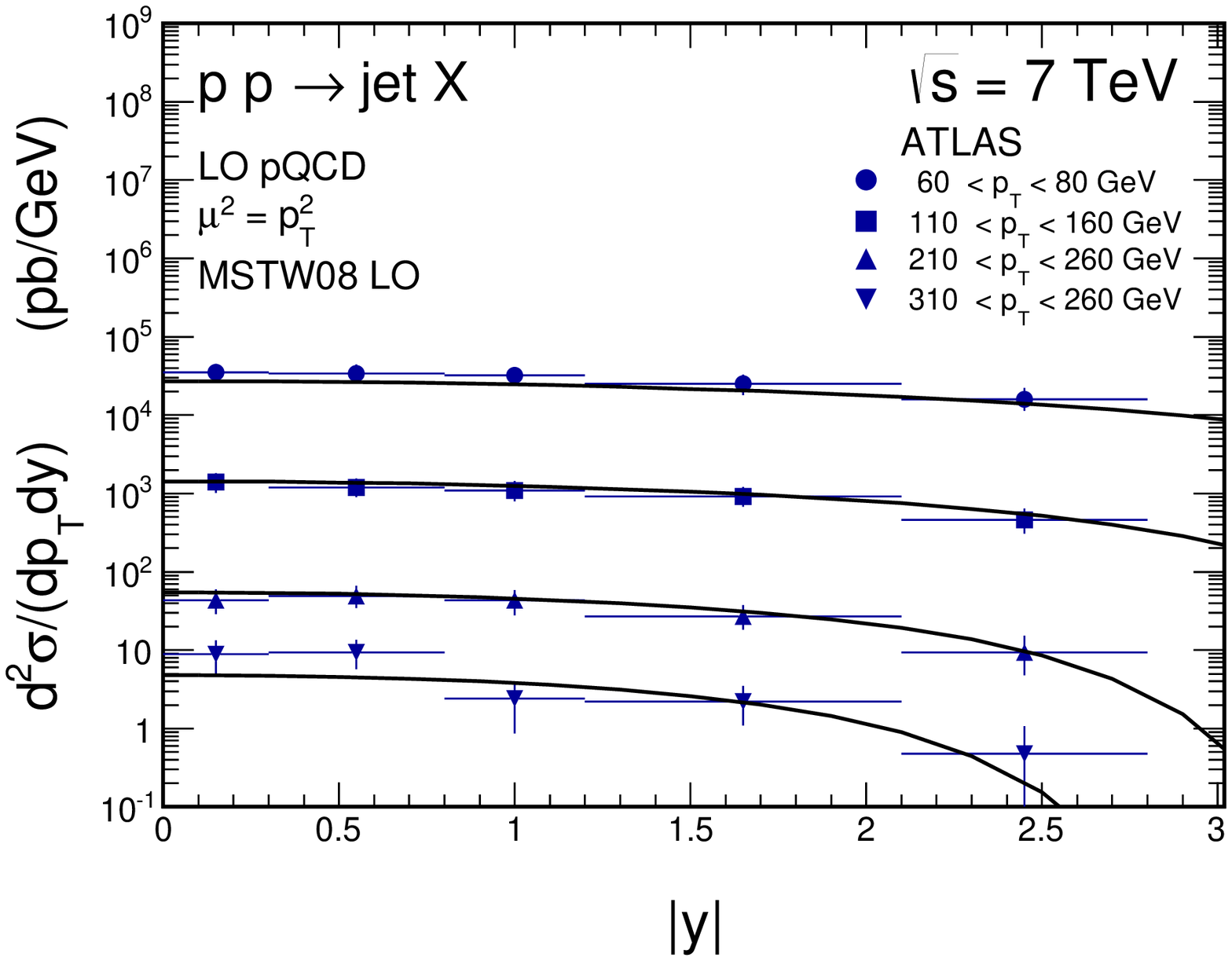}}
\end{minipage}
   \caption{
\small Transverse momentum distribution of jets for different regions
of the jet rapidity (left panel) and corresponding rapidity distribution
of jets with different cuts in $p_t$ as specified in the figure
caption of the right panel. The theoretical calculations were
performed with the MSTW08 set of parton distributions \cite{MSTW08}.
The data points were obtained by the ATLAS collaboration \cite{ATLASjets}.
}
 \label{fig:pt-and-y-spectra-ATLASjets}
\end{figure}

Before we shall show our results for rapidity-distant-jet correlations 
we wish to verify the quality of description of observables 
for inclusive jet production. In Fig.~\ref{fig:pt-and-y-spectra-ATLASjets}
we show distributions in the jet transverse momentum for different
intervals of jet (pseudo)rapidity (left panel) and distribution in jet (pseudo)rapidity for different intervals of jet transverse momentum (right panel).
In this calculations we have used
the MSTW08 PDFs \cite{MSTW08}. The agreement with recent ATLAS data
\cite{ATLASjets} is fairly reasonable which alows us
to use the same distributions for first evalution of the DPS effects 
for large rapidity distances between jets. 

\begin{figure}[!h]
\begin{minipage}{0.47\textwidth}
 \centerline{\includegraphics[width=1.0\textwidth]{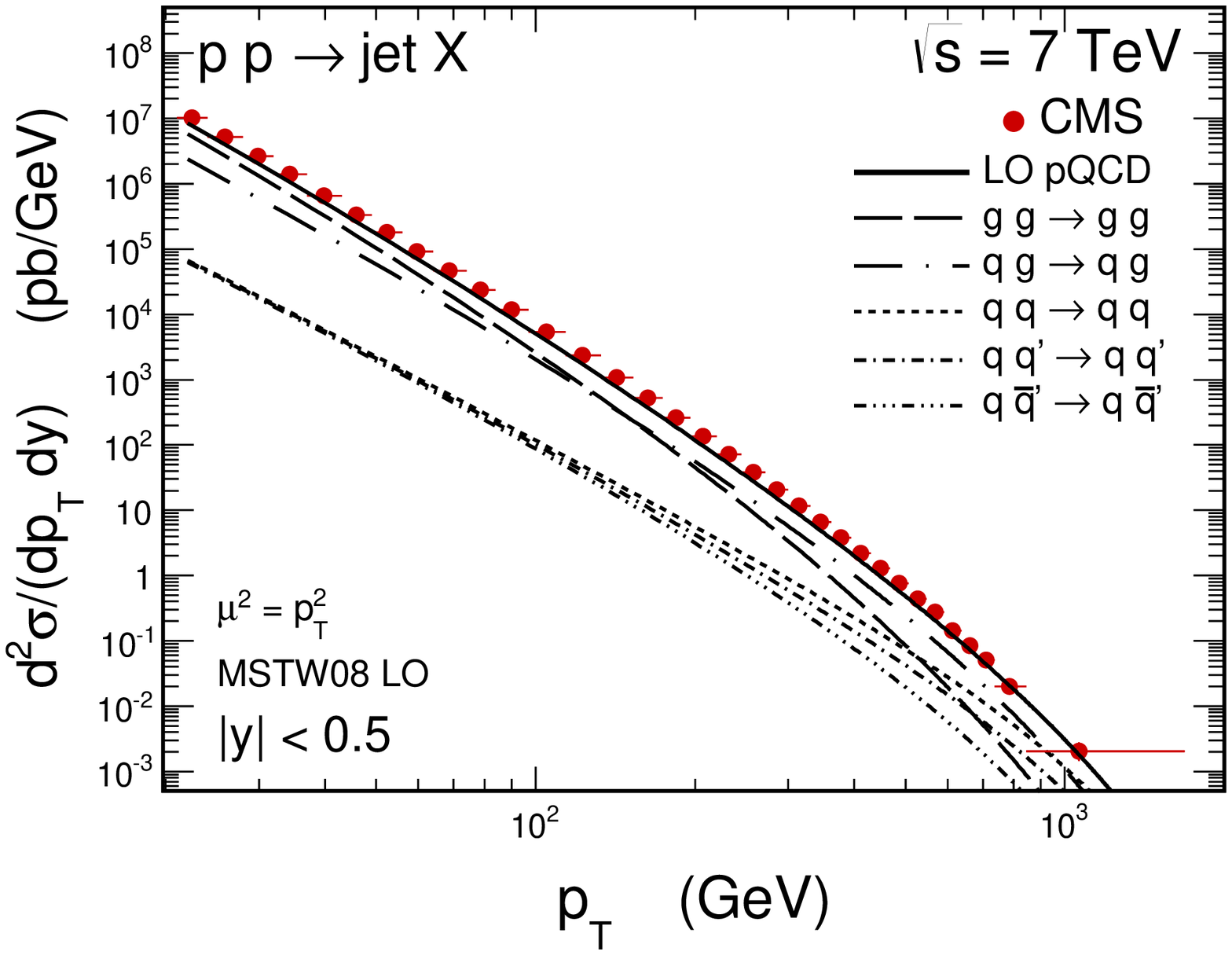}}
\end{minipage}
\hspace{0.5cm}
\begin{minipage}{0.47\textwidth}
 \centerline{\includegraphics[width=1.0\textwidth]{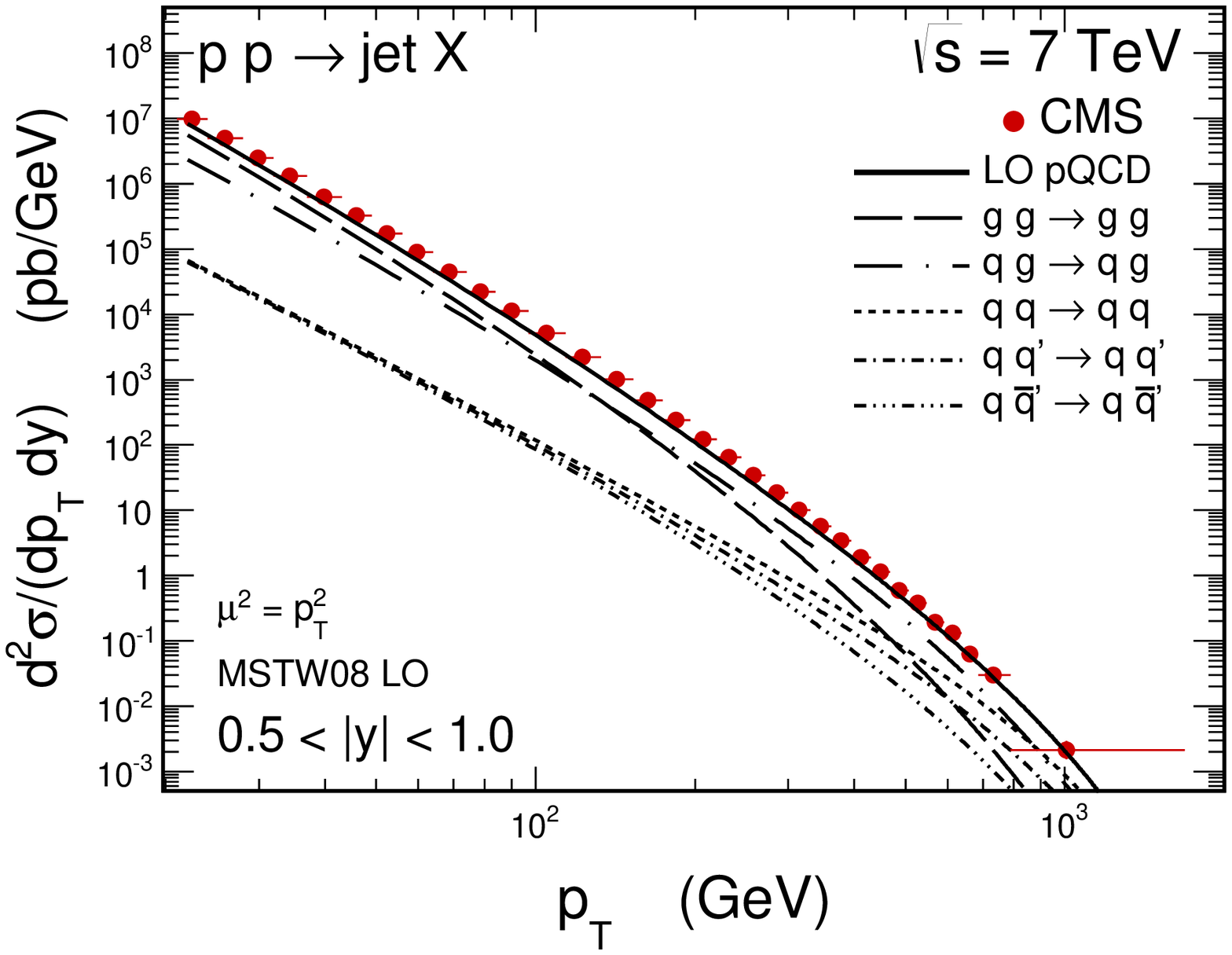}}
\end{minipage}\\
\begin{minipage}{0.47\textwidth}
 \centerline{\includegraphics[width=1.0\textwidth]{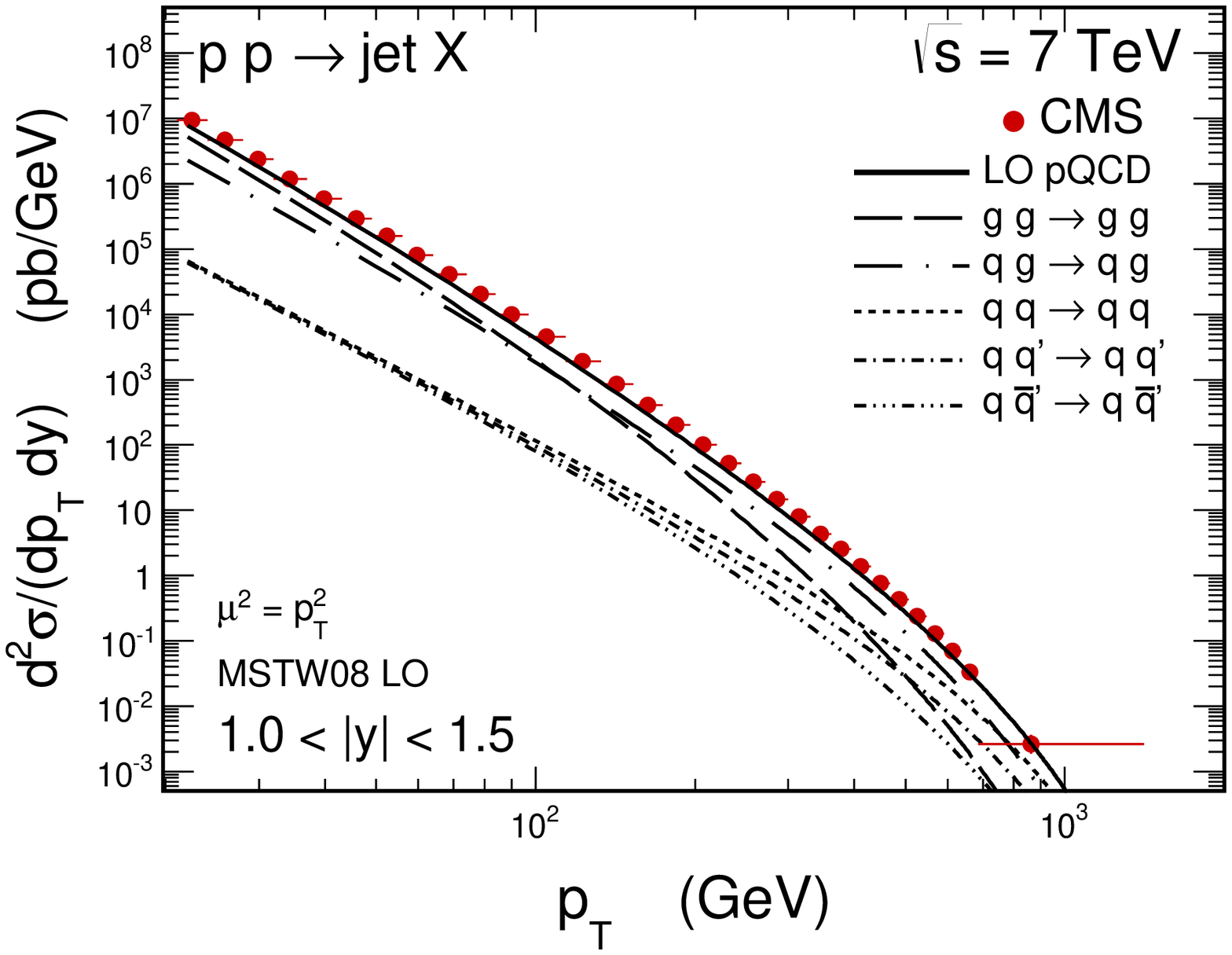}}
\end{minipage}
\hspace{0.5cm}
\begin{minipage}{0.47\textwidth}
 \centerline{\includegraphics[width=1.0\textwidth]{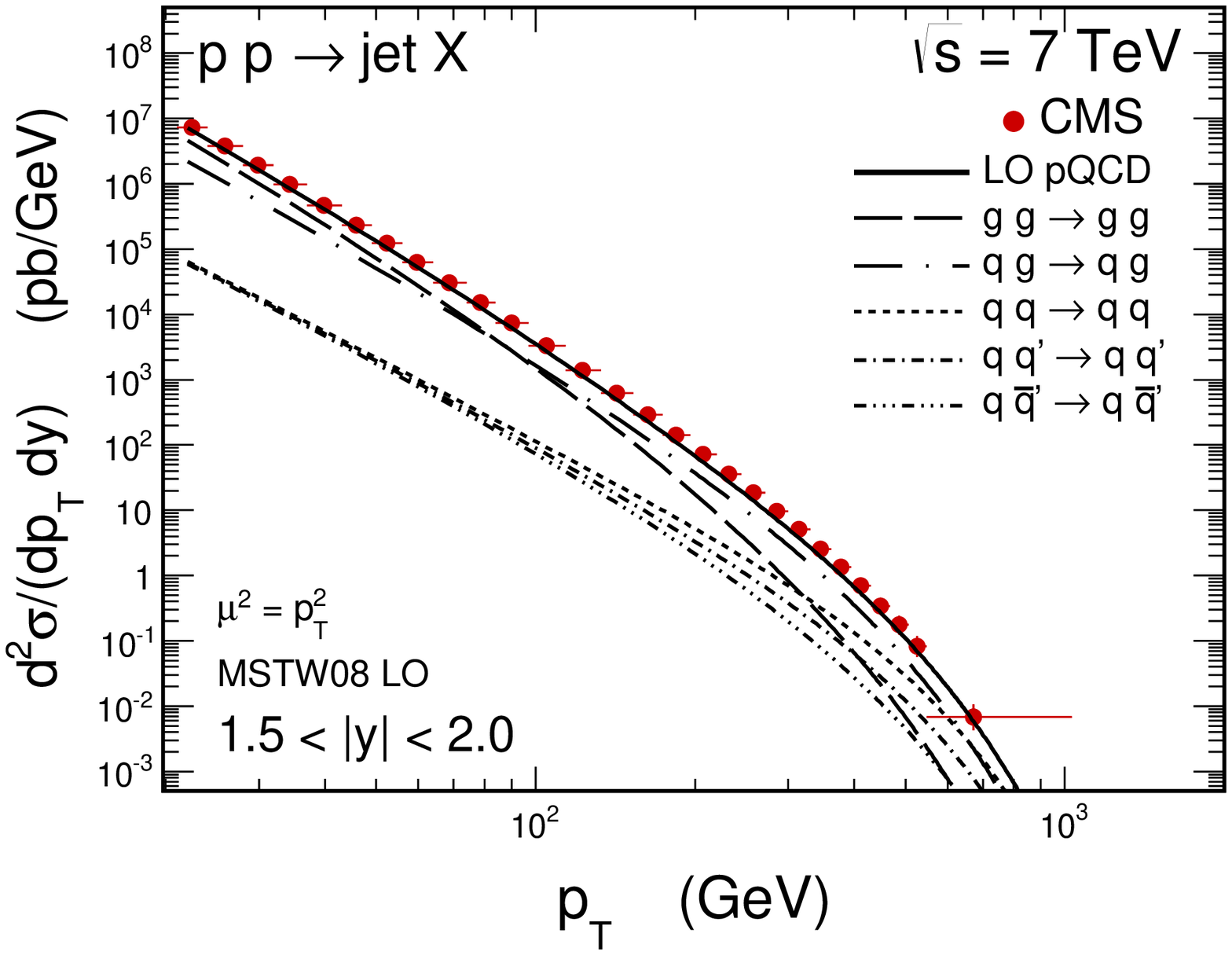}}
\end{minipage}\\
\begin{minipage}{0.47\textwidth}
 \centerline{\includegraphics[width=1.0\textwidth]{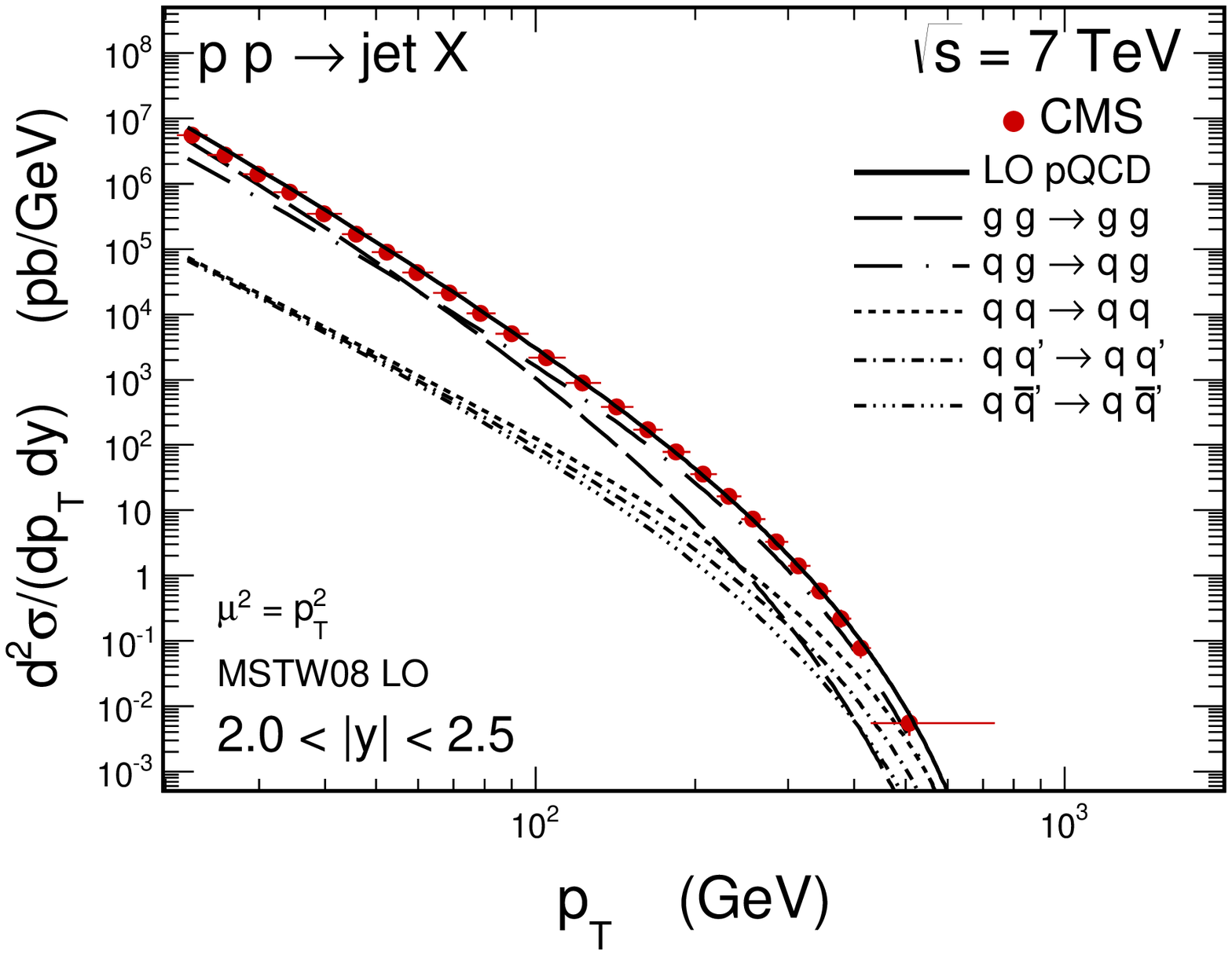}}
\end{minipage}
\hspace{0.5cm}
\begin{minipage}{0.47\textwidth}
 \centerline{\includegraphics[width=1.0\textwidth]{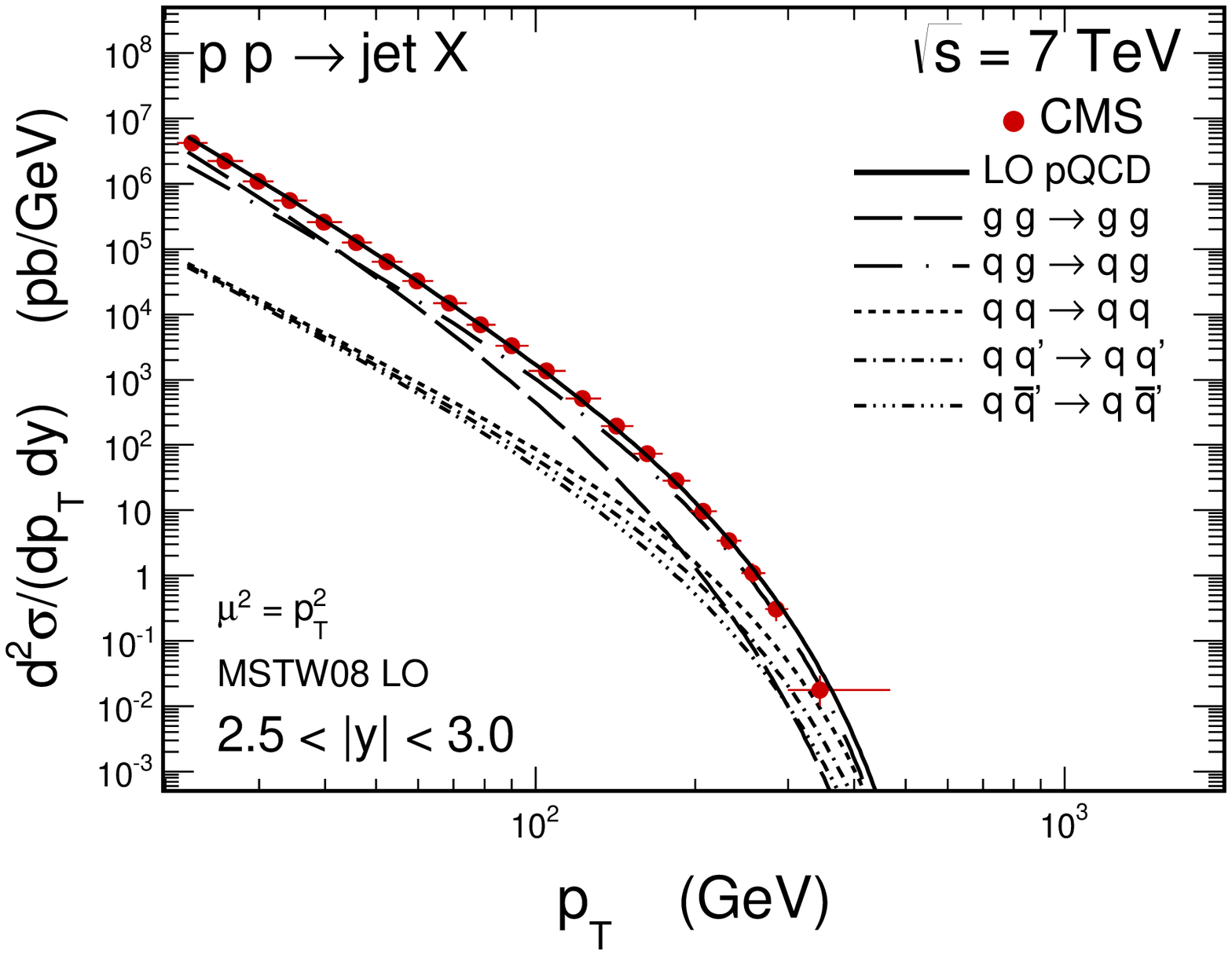}}
\end{minipage}

   \caption{
\small The same as in the previous figure but now together with the CMS
experimental data \cite{CMSjets}.
In addition, we show decomposition into different partonic components 
as explained in the figure caption.
}
 \label{fig:pt-and-y-spectra-CMSjets}
\end{figure}

In Fig.~\ref{fig:pt-and-y-spectra-CMSjets} we compare our calculation with 
the CMS data \cite{CMSjets}. In addition, we show contributions of different
partonic mechanisms. In all rapidity intervals the gluon-gluon and quark-gluon
(gluon-quark) contributions clearly dominate over the other
contributions and in practice it is sufficient to include only
these subprocesses in further analysis.

\begin{figure}[!h]
\begin{minipage}{0.47\textwidth}
 \centerline{\includegraphics[width=1.0\textwidth]{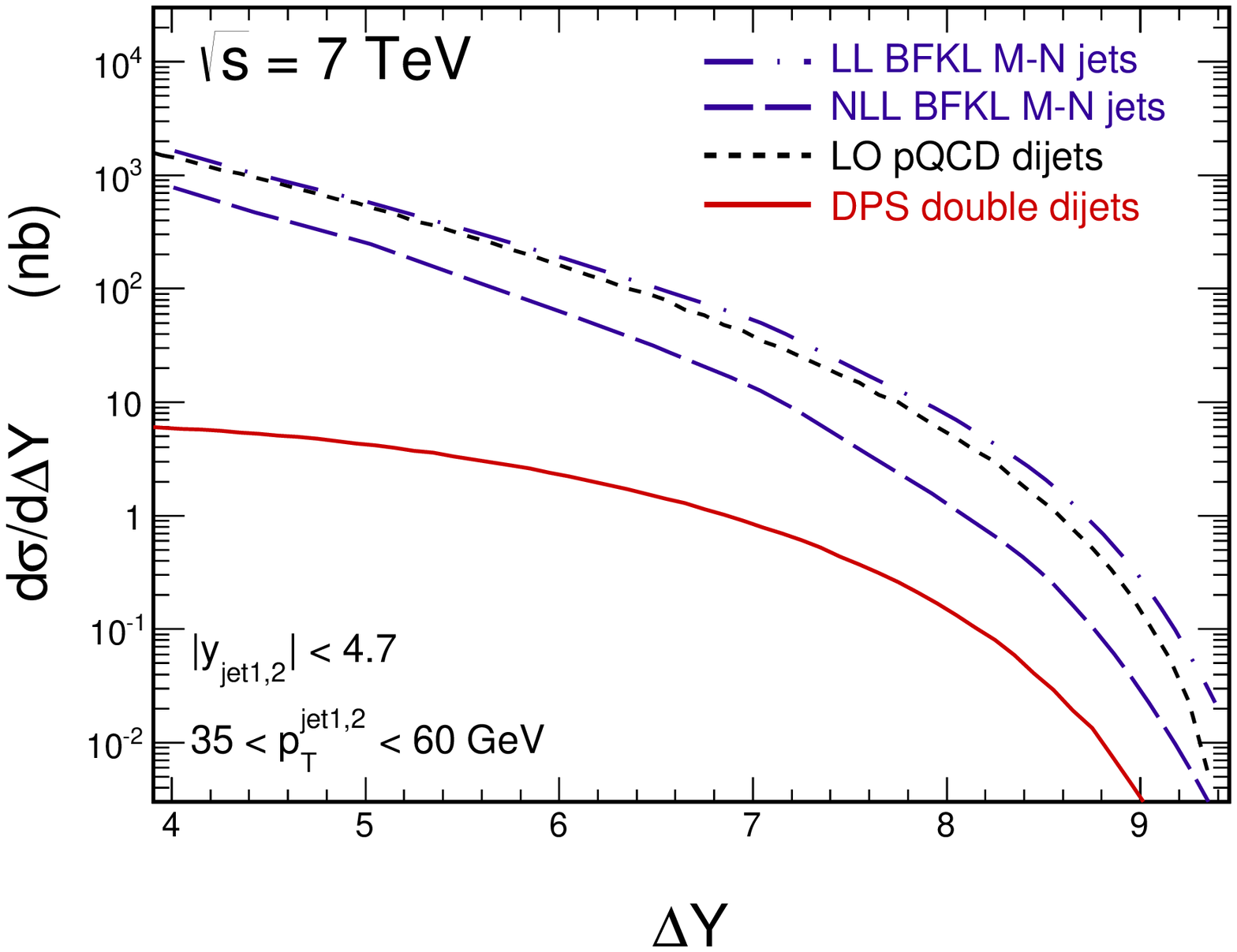}}
\end{minipage}
\hspace{0.5cm}
\begin{minipage}{0.47\textwidth}
 \centerline{\includegraphics[width=1.0\textwidth]{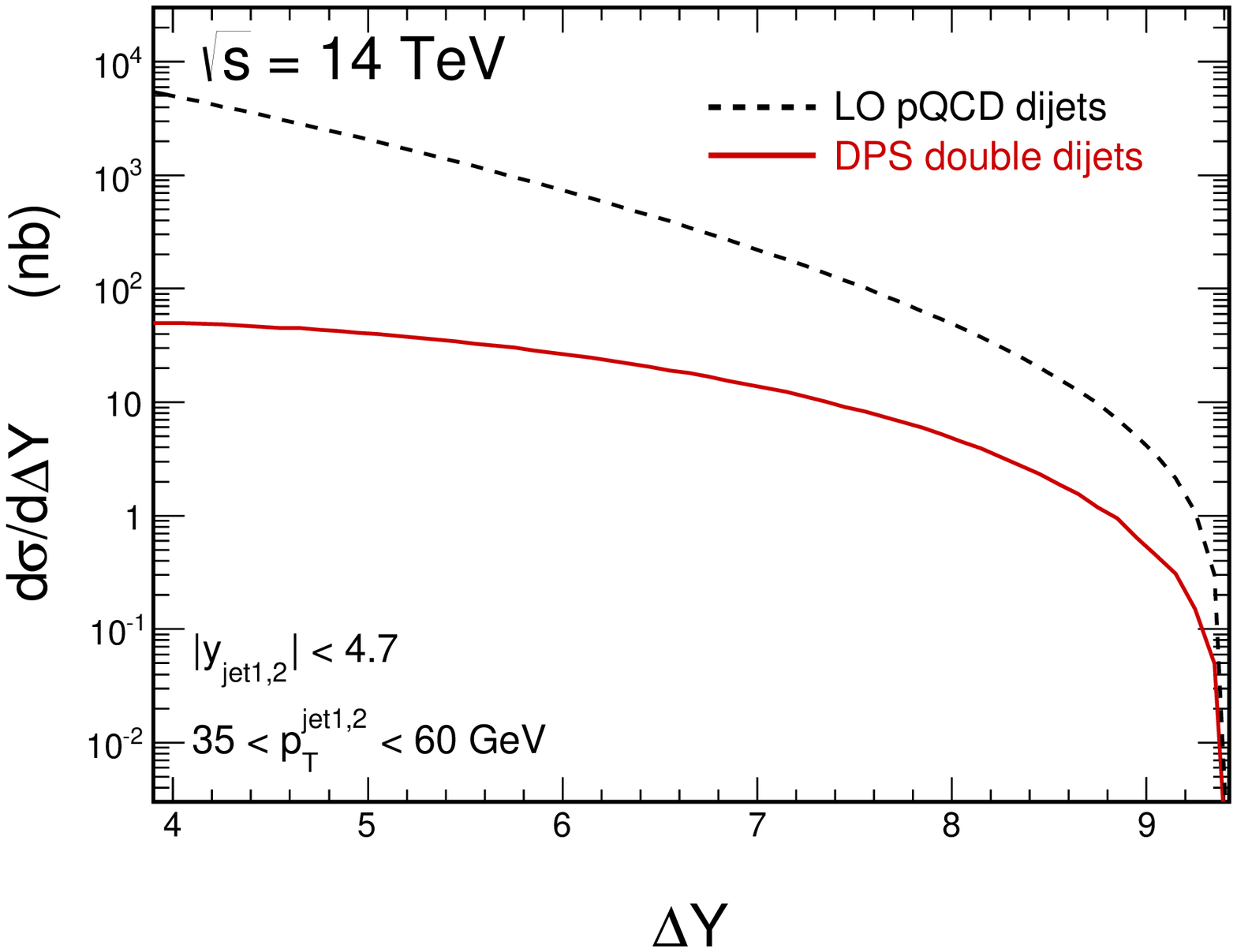}}
\end{minipage}
   \caption{
\small Distribution in rapidity distance between jets 
(35 GeV $< p_t <$ 60 GeV) 
with maximal (the most positive) and minimal (the most negative)
rapidities. The collinear pQCD result is shown by the short-dashed line
and the DPS result by the solid line for $\sqrt{s}$ = 7 TeV (left panel)
and $\sqrt{s}$ = 14 TeV (right panel). For comparison we show also
results for the classical BFKL Mueller-Navelet jets in leading-logarithm 
and next-to-leading-order logarithm approaches from Ref.~\cite{Ducloue:2013hia}.
}
 \label{fig:Deltay1}
\end{figure}

Now we shall proceed to the jets with large rapidity separation.
In Fig.~\ref{fig:Deltay1} we show distribution in the rapidity 
distance between two jets in leading-order collinear calculation
and between the most distant jets in rapidity in the case of four DPS jets.
In this calculation we have included cuts characteristic for the
CMS expriment \cite{CMS_private}: $y_1, y_2 \in$ (-4.7,4.7),
$p_{1t}, p_{2t} \in$ (35 GeV, 60 GeV).
For comparison we show also results for the LL and NLL BFKL calculation for MN jet from
Ref.~\cite{Ducloue:2013hia}. For this kinematics the DPS jets
give sizeable relative contribution only at large rapidity distance.
However, the four-jet (DPS) and dijet (LO SPS) final state can 
be easily distinguished and, in principle, one can concentrate on the
DPS contribution which is interesting by itself.
The NLL BFKL cross section (long-dashed line) is smaller than that for 
the LO collinear approach (short-dashed line).

\begin{figure}[!h]
\begin{minipage}{0.47\textwidth}
 \centerline{\includegraphics[width=1.0\textwidth]{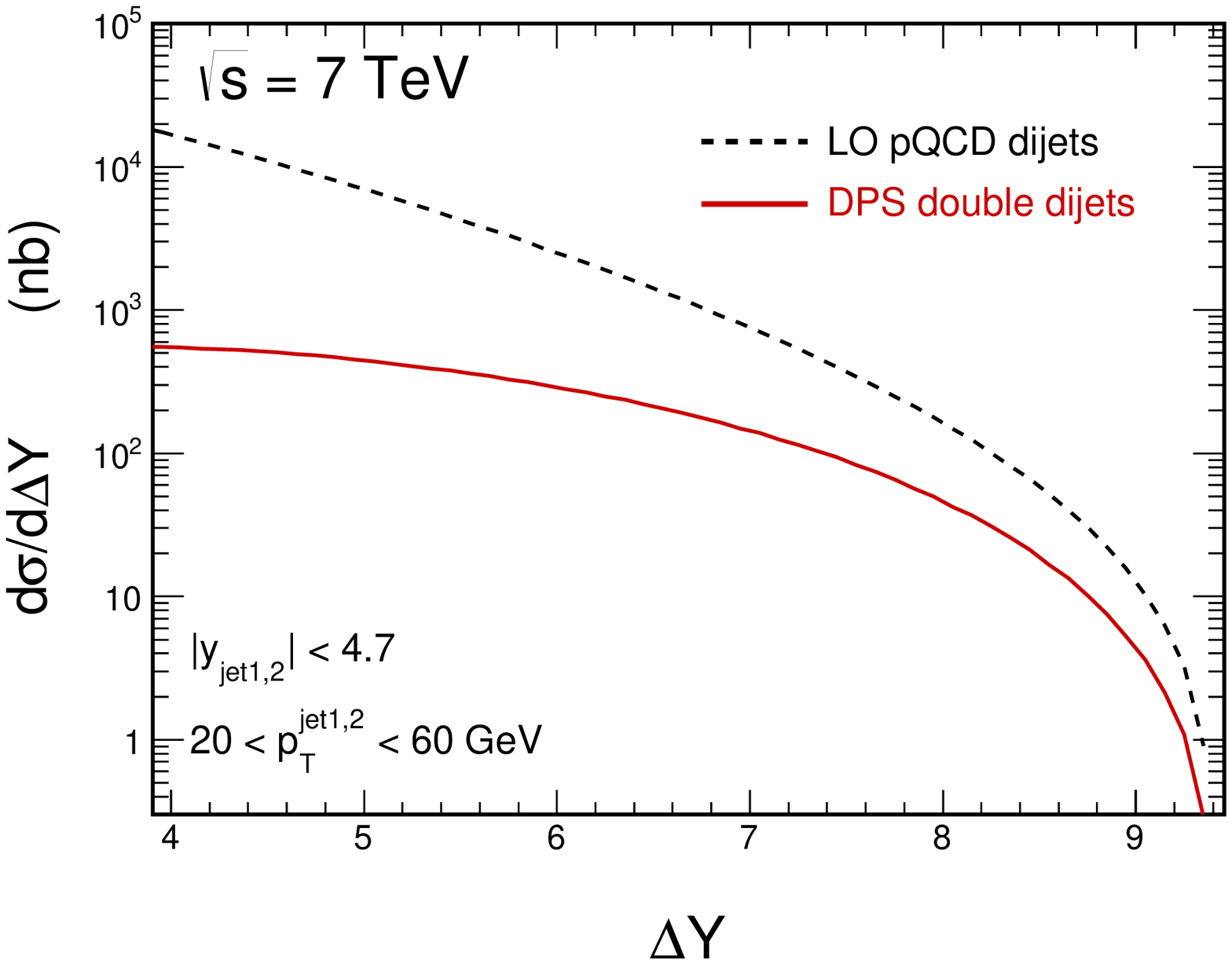}}
\end{minipage}
\hspace{0.5cm}
\begin{minipage}{0.47\textwidth}
 \centerline{\includegraphics[width=1.0\textwidth]{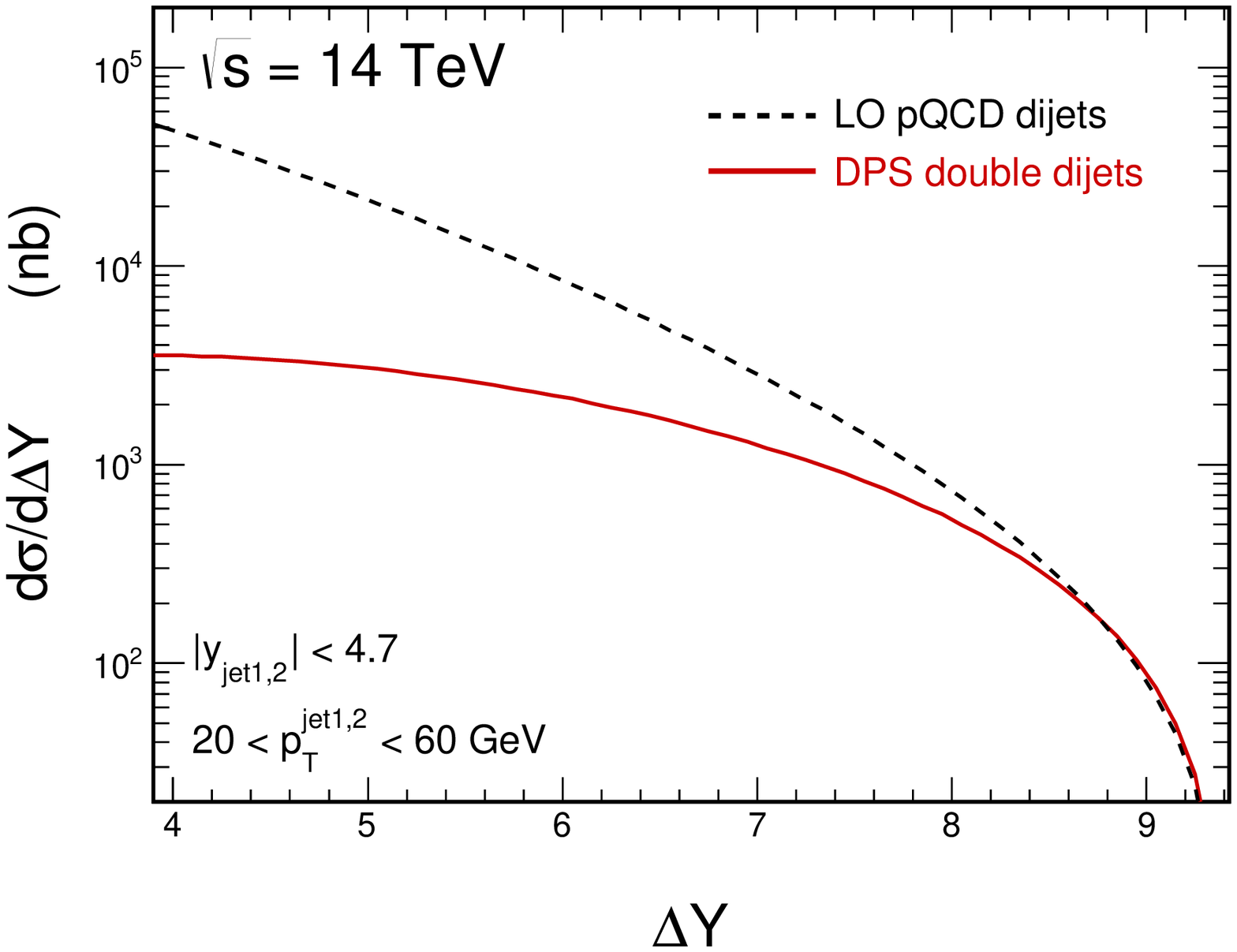}}
\end{minipage}\\
\begin{minipage}{0.47\textwidth}
 \centerline{\includegraphics[width=1.0\textwidth]{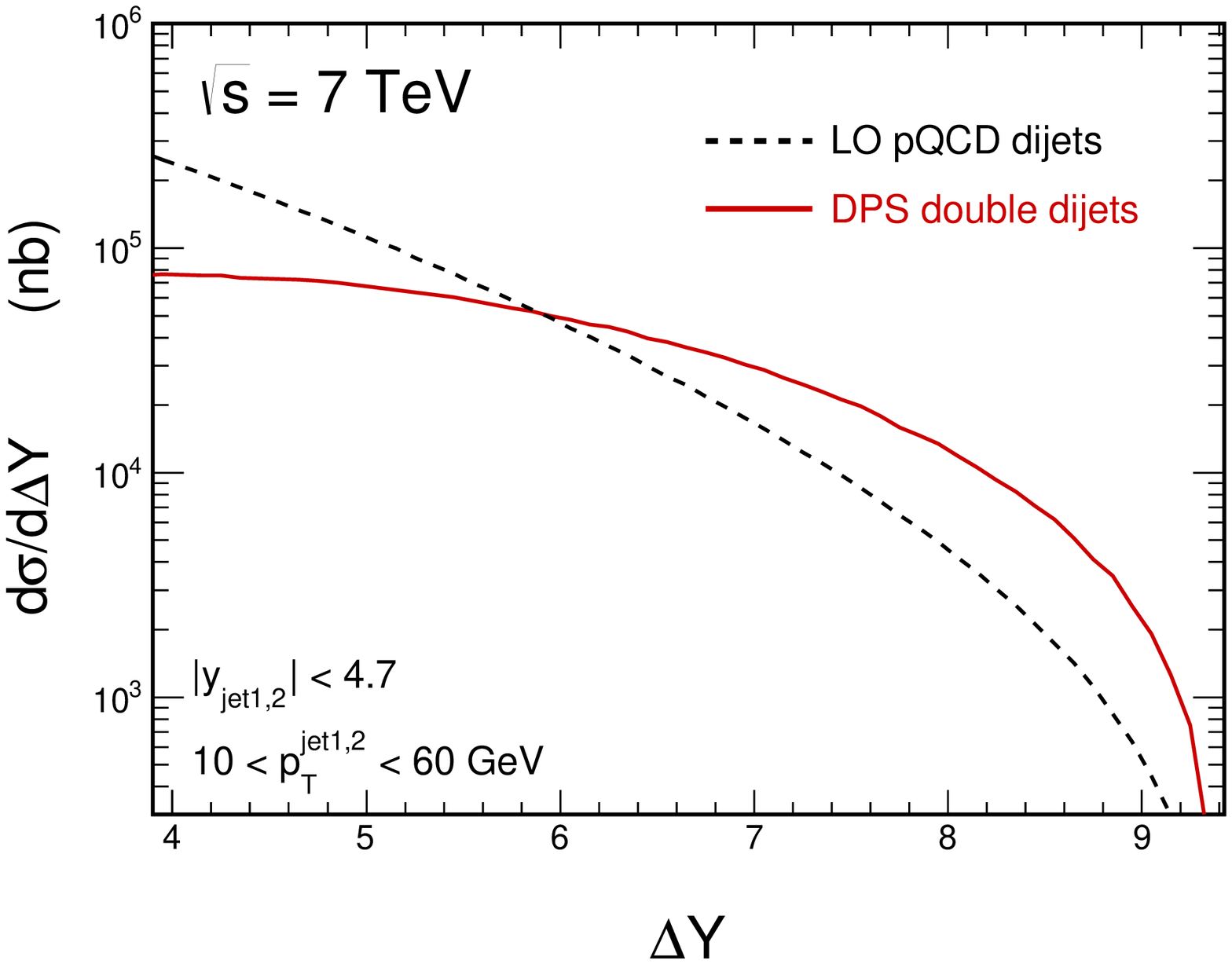}}
\end{minipage}
\hspace{0.5cm}
\begin{minipage}{0.47\textwidth}
 \centerline{\includegraphics[width=1.0\textwidth]{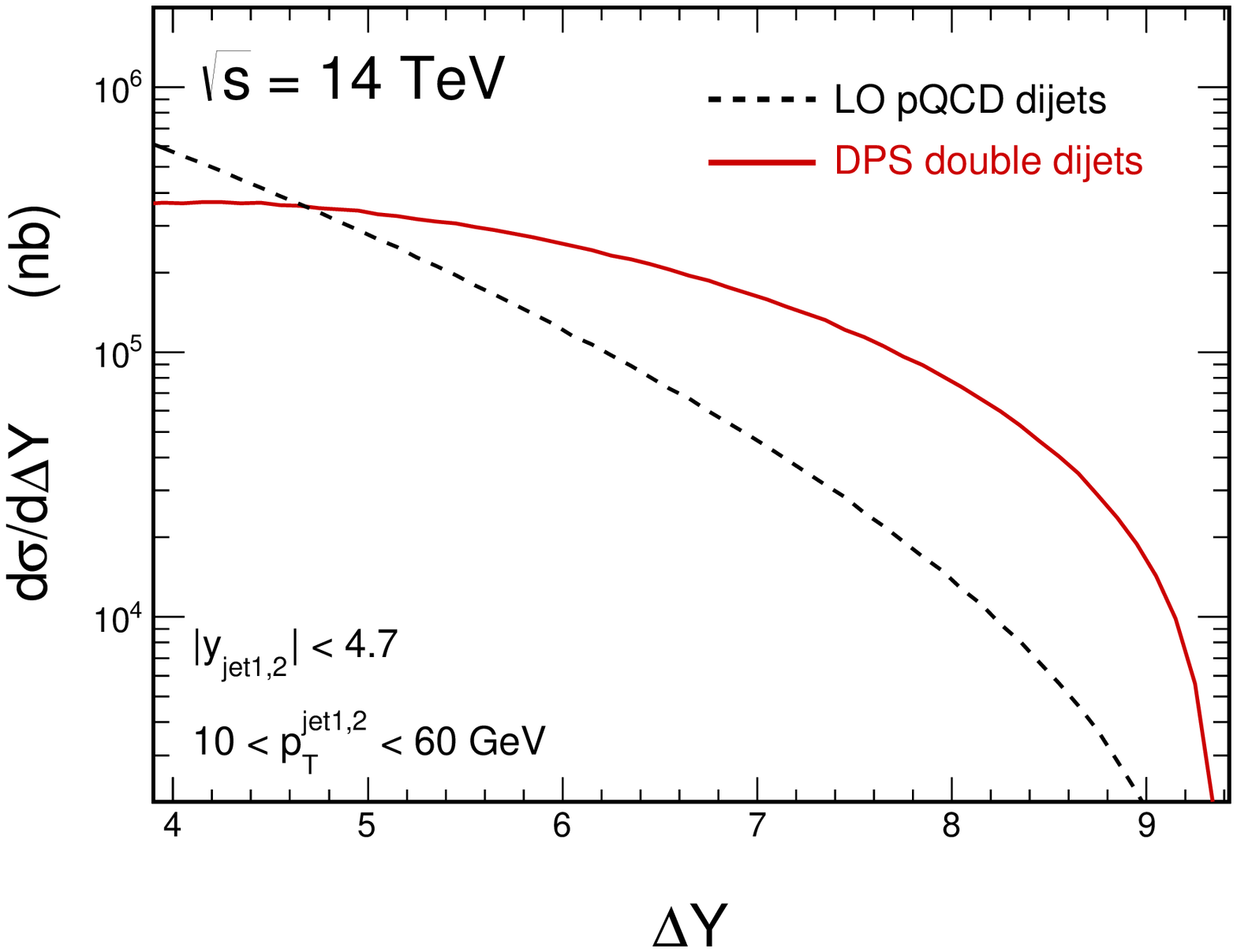}}
\end{minipage}
   \caption{
\small The same as in the previous figure but now for somewhat smaller lower
cut on minijet transverse momentum.
}
 \label{fig:Deltay-2}
\end{figure}

\begin{figure}[!h]
\begin{minipage}{0.47\textwidth}
 \centerline{\includegraphics[width=1.0\textwidth]{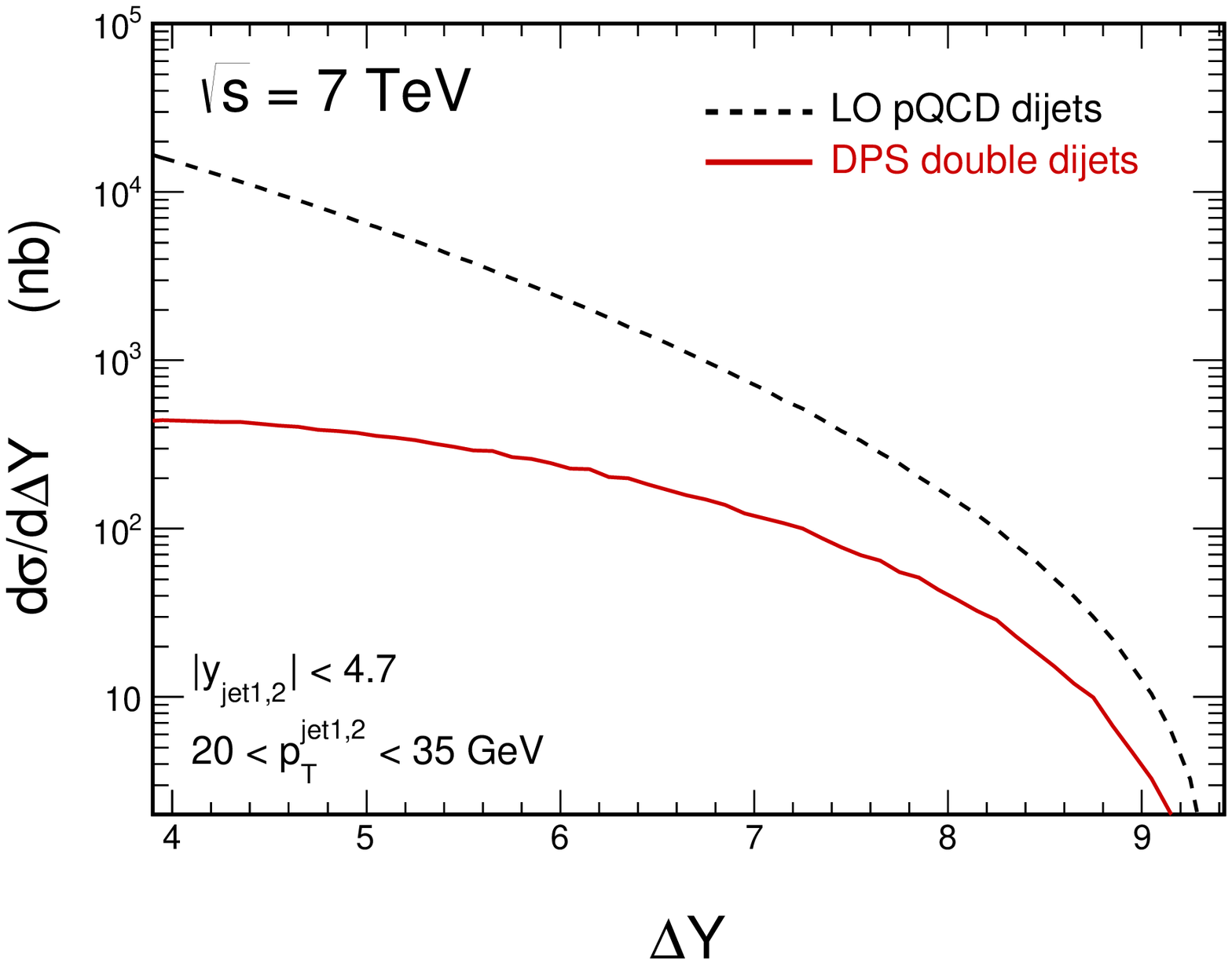}}
\end{minipage}
\hspace{0.5cm}
\begin{minipage}{0.47\textwidth}
 \centerline{\includegraphics[width=1.0\textwidth]{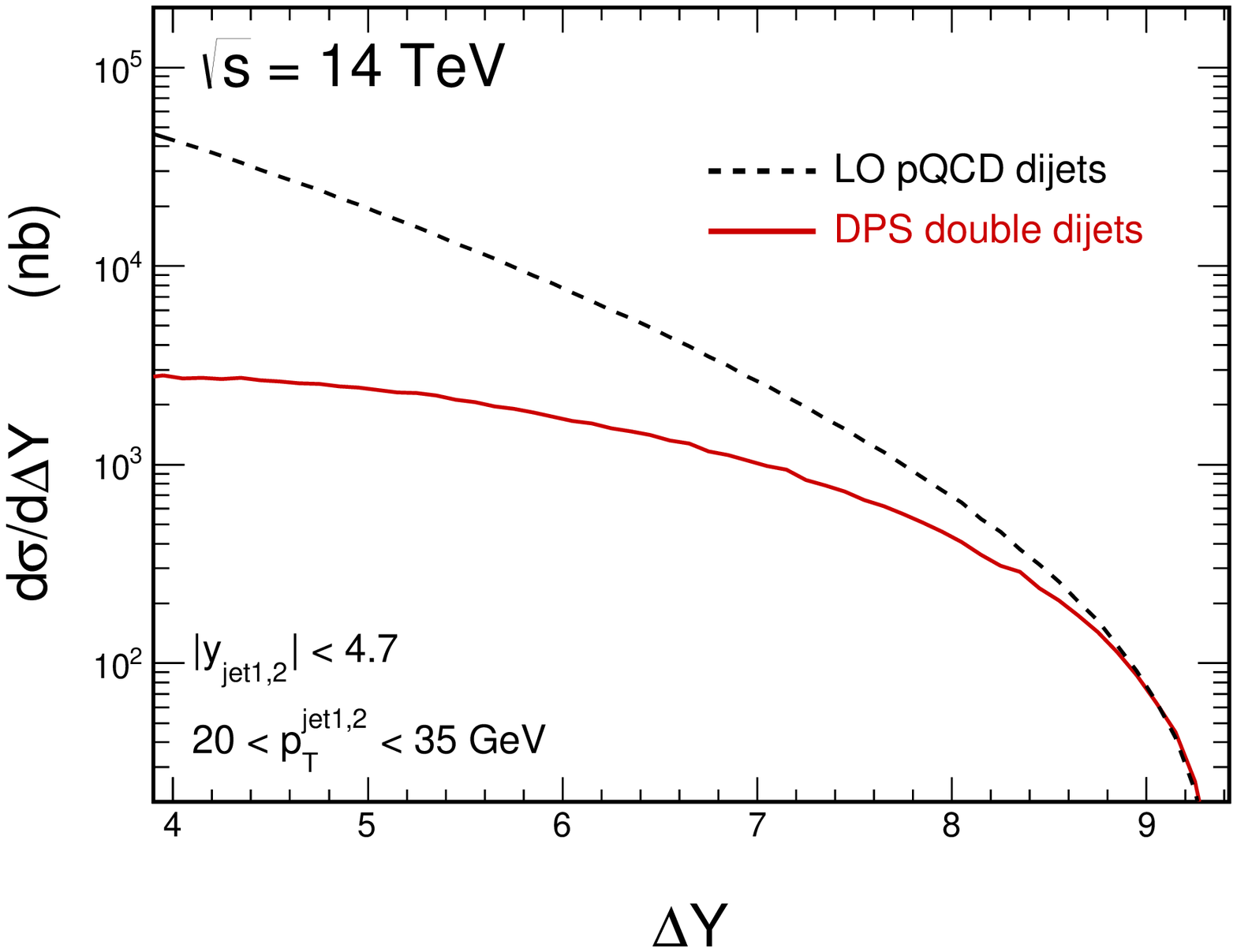}}
\end{minipage}\\
\begin{minipage}{0.47\textwidth}
 \centerline{\includegraphics[width=1.0\textwidth]{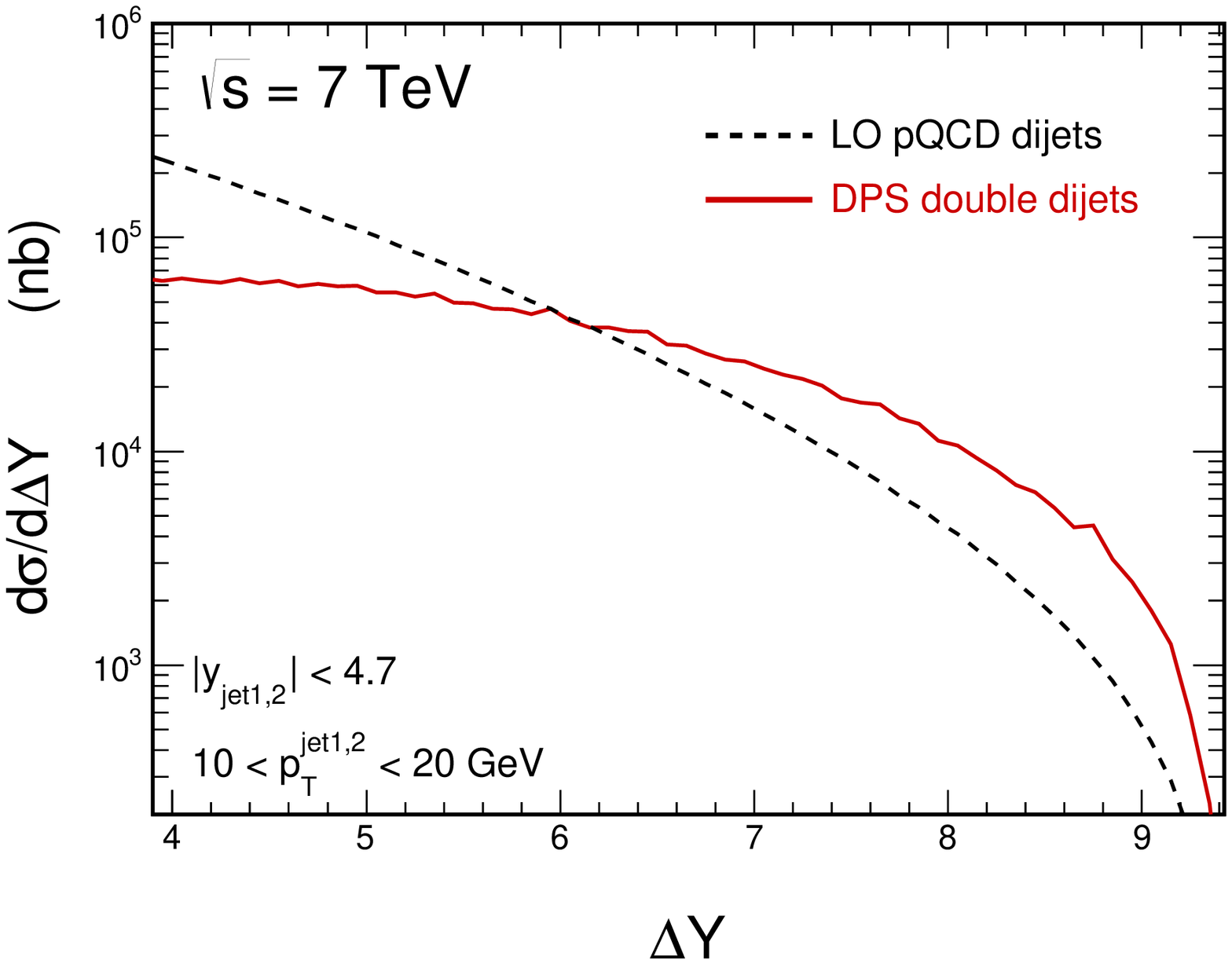}}
\end{minipage}
\hspace{0.5cm}
\begin{minipage}{0.47\textwidth}
 \centerline{\includegraphics[width=1.0\textwidth]{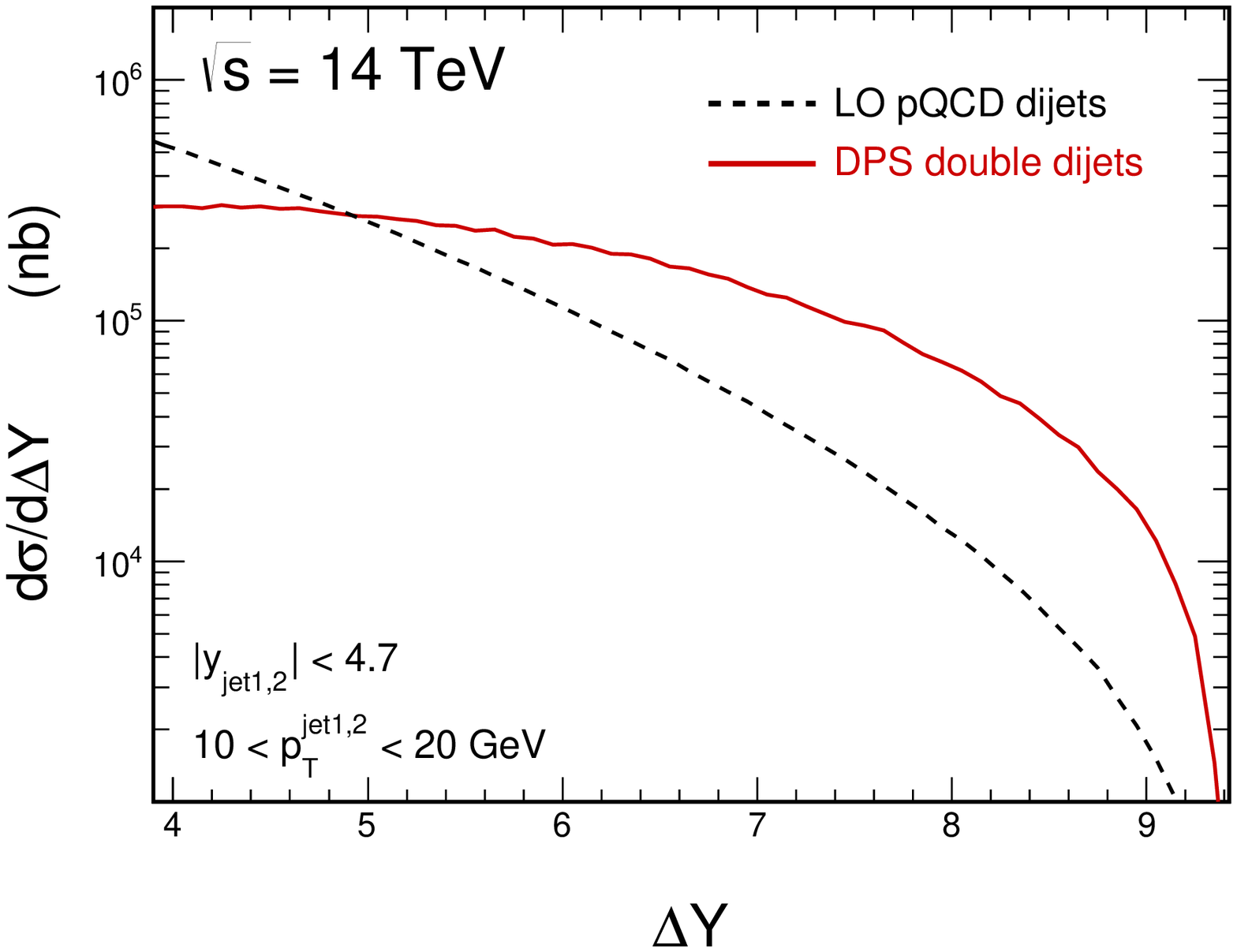}}
\end{minipage}
   \caption{
\small The same as in the previous figures but now for somewhat smaller upper
cut on minijet transverse momentum in addition.
}
 \label{fig:Deltay-3}
\end{figure}

As for the BFKL Mueller-Navelet jets the DPS contribution is growing with 
deacreasing jet transverse momenta.
Therefore let us now discuss results for even smaller transverse momenta.
In Fig.~\ref{fig:Deltay-2} we show rapidity-distance
distribution for even smaller lowest transverse momentum of 
the "jet". A measurement of such minijets may be, however, difficult. 
Now the DPS contribution may even exceed the standard SPS 
dijet contribution, especially at the nominal LHC energy. 
In Fig.~\ref{fig:Deltay-3} we lower
in addition the upper limit for the jets. The situation does not improve further.
How to measure such (mini)jets is an open issue. In principle,
one could measure for instance correlations of 
semihard ($p_t \sim$ 10 GeV) neutral pions with the help of 
so-called zero-degree calorimeters (ZDC) which are installed by
all major LHC experiments.
Other possibilities could be considered too.

\begin{figure}[!h]
\begin{minipage}{0.47\textwidth}
 \centerline{\includegraphics[width=1.0\textwidth]{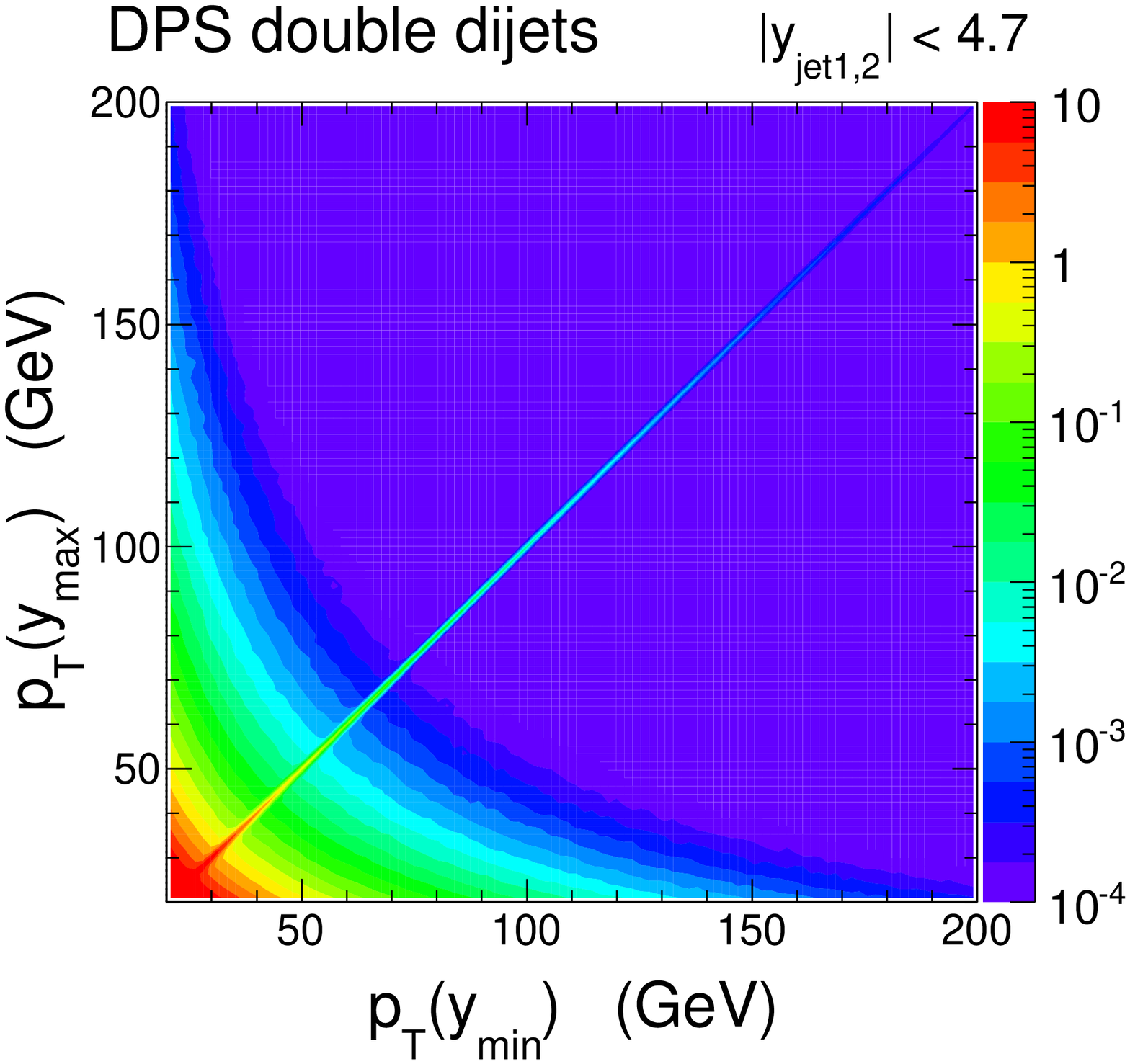}}
\end{minipage}
\hspace{0.5cm}
\begin{minipage}{0.47\textwidth}
 \centerline{\includegraphics[width=1.0\textwidth]{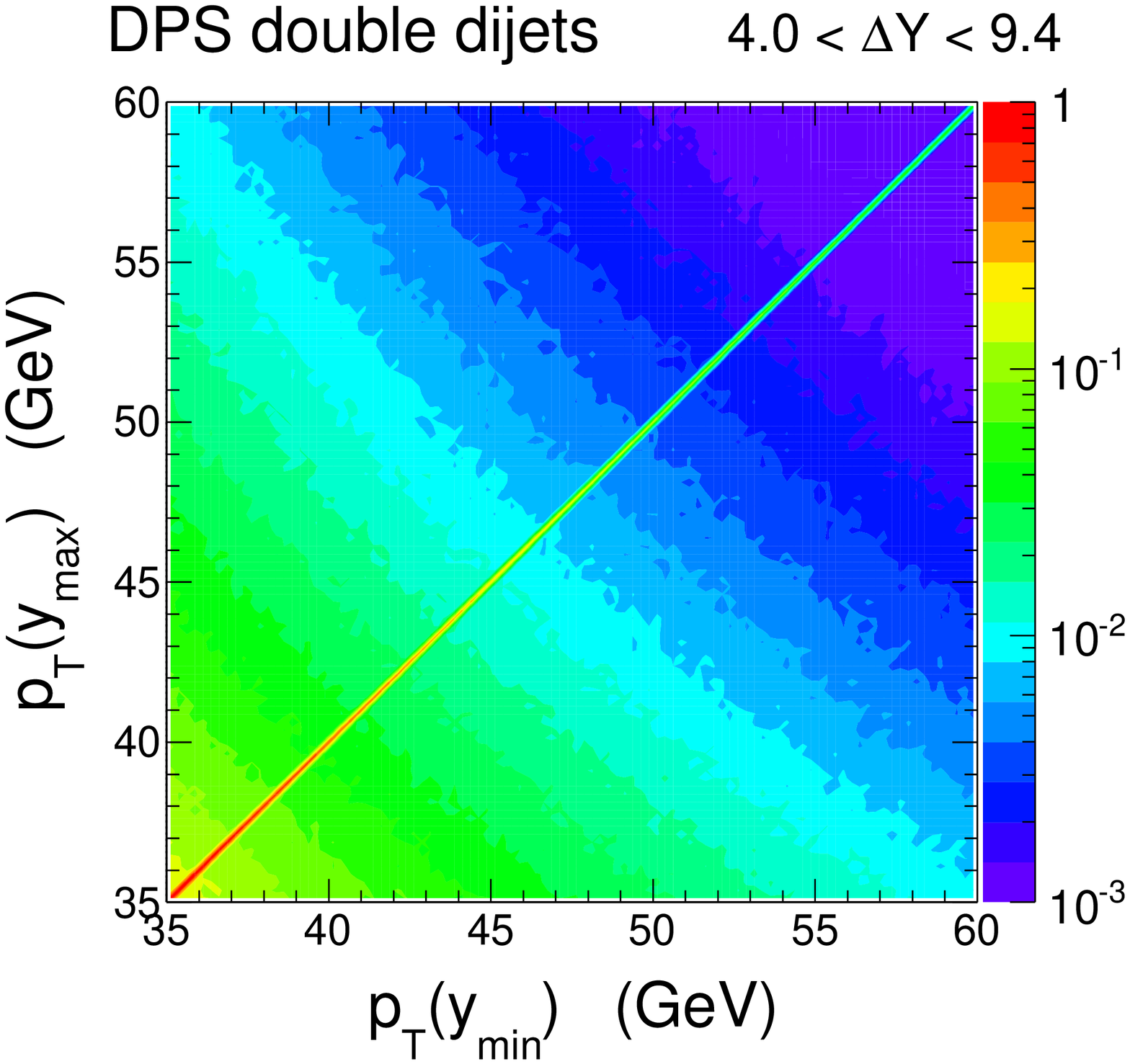}}
\end{minipage}
   \caption{
\small The double-differential distribution in transverse momenta
of jet with minimal ($p_T(y_{min})$)
and maximal ($p_T(y_{max})$) rapidities. In the left panel
we show the result for the full range of rapidities and in the right
panel only for large rapidity separations as defined previously.
}
 \label{fig:pTmin-pTmax}
\end{figure}

\begin{figure}[!h]
\begin{minipage}{0.47\textwidth}
 \centerline{\includegraphics[width=1.0\textwidth]{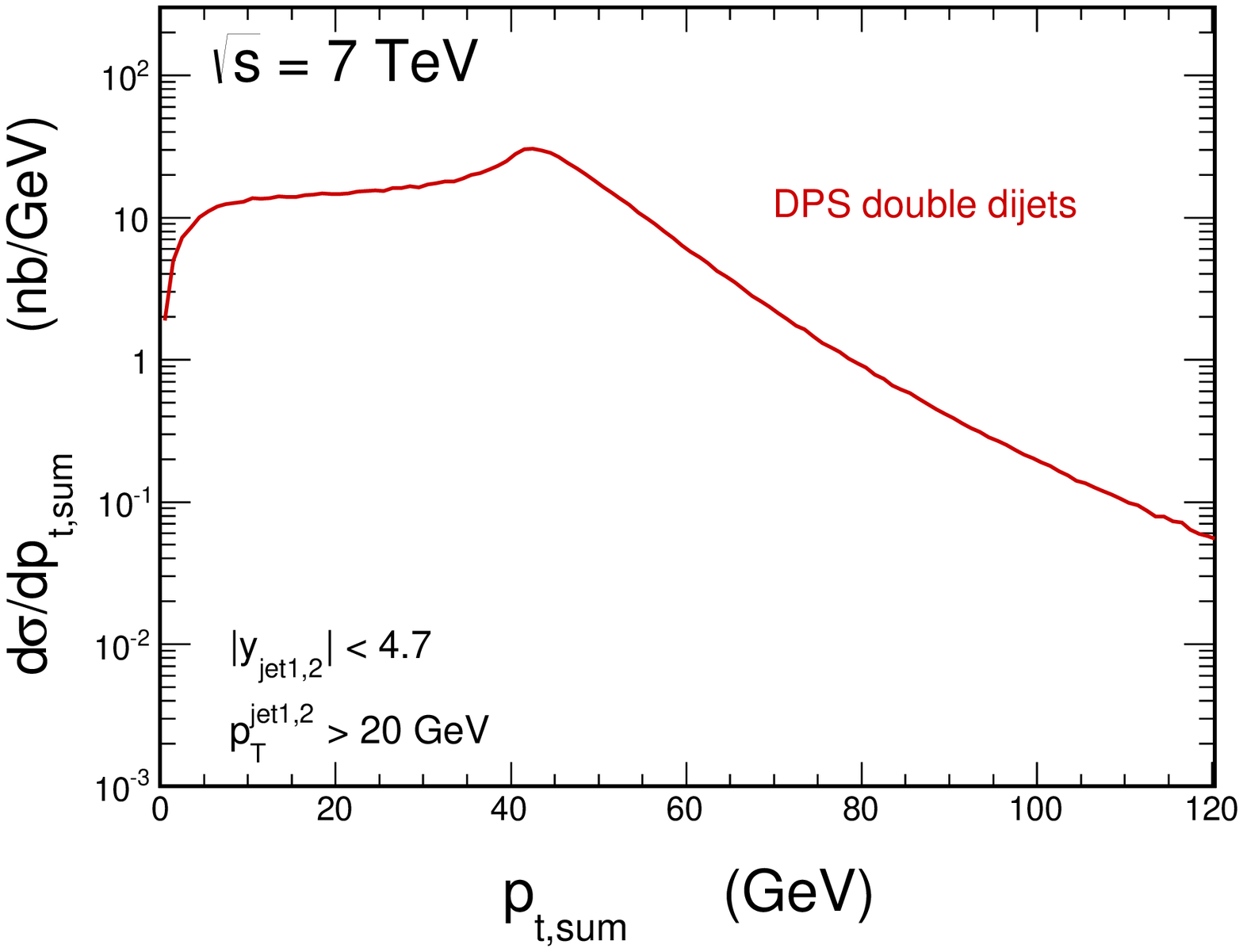}}
\end{minipage}
\hspace{0.5cm}
\begin{minipage}{0.47\textwidth}
 \centerline{\includegraphics[width=1.0\textwidth]{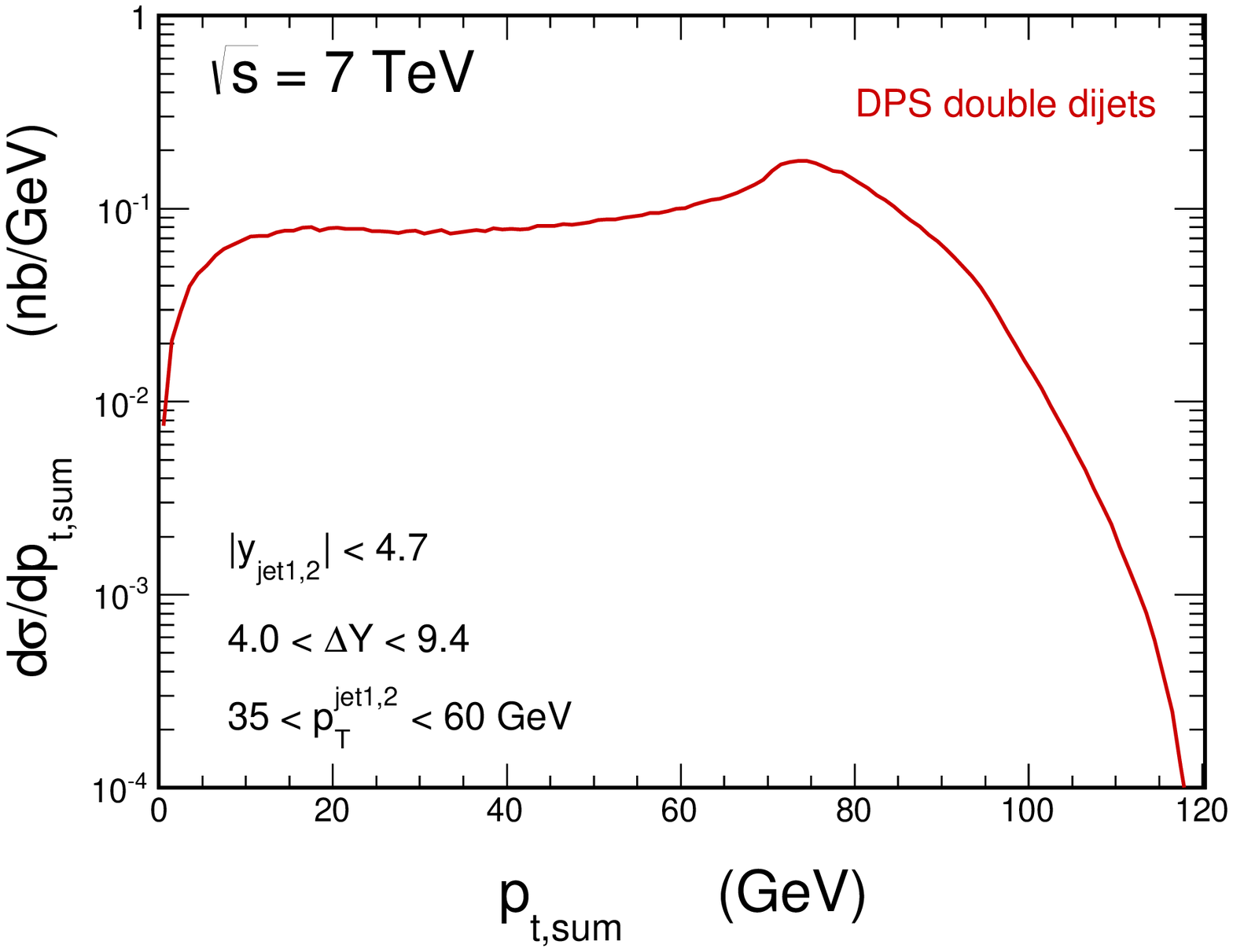}}
\end{minipage}
   \caption{
\small Distribution in transverse momentum imbalance 
(vector sum of transverse momenta of jets) 
between the jets with "minimal" and "maximal" rapidities.
}
 \label{fig:pTsum-1}
\end{figure}

Now we wish to concentrate ourselves on correlations between transverse
momenta of the rapidity-distant jets. In our case the
large-rapidity-distance jets are coming from different partonic
scatterings and are therefore quite uncorrelated. 
In Fig.~\ref{fig:pTmin-pTmax} we present our results. 
The ($p_{1t},p_{2t}$) distribution for the DPS mechanism is rather
different than similar distributions for dijet SPS \cite{Rybarska}
and MN jets \cite{DelDuca:1994ng}.
The dijets from the SPS as well as jets from the same partonic scattering in DPS are correlated along the $p_{1t} = p_{2t}$ 
diagonal \cite{Rybarska} (see straight diagonal line in Fig.~\ref{fig:pTmin-pTmax}).
In principle, one could eliminate this region by dedicated cuts.
How the situation looks in the BFKL calculation can be already seen
from simple LL calculation \cite{DelDuca:1994ng}.
The CMS collaboration could make such two-dimensional studies.
Another alternative are studies of distributions in the transverse 
momentum imbalance $\vec{p}_{t,sum} = \vec{p}_{1t} + \vec{p}_{2t}$
between the rapidity-distant jets. In Fig.~\ref{fig:pTsum-1}
we show distributions for full range of rapidity distances (left panel)
as well as for large-rapidity-distance jets (right panel).
The DPS mechanism generates situations with large transverse momentum
imbalance. This could be used in addition to enhance the content
of DPS effects by taking a lower cut on the dijet imbalance.
The transverse momentum imbalance for jets remote in rapidity
is bigger than that for jets close in rapidity.
The corresponding distribution for Mueller-Navelet jets has maximum at $p_{t,sum} \sim$ 0.
It would be interesting to calculate the transverse momentum imbalance
also for SPS dijets as well as for the Mueller-Navelet jets.
This clearly goes beyond the scope of this short note.

Finally we wish to discuss azimuthal correlations between the jets distant
in rapidity.
The azimuthal angle distributions for the Mueller-Navelet jets were
calculated by many groups and we will not repeat such calculations here.
The DPS jets are fully uncorrelated, at least in our approach.
This is expected to be different for the SPS dijets 
(delta function $\delta(\phi - \pi)$ in the
leading-order collinear approach) as well as for the classical 
Mueller-Navelet jets. The SPS dijet azimuthal correlations 
as well as the transverse momentum imbalance distribution could be
easily calculated in the $k_t$-factorization approach \cite{Leonidov:1999nc,Ostrovsky:1999kj,Bartels:2006hg,Rybarska,Nefedov:2013ywa}. In this approach one avoids singularities
present in the fixed-order collinear approach.

A contamination of the large-rapidity-distance jets by the DPS effects 
may distort the information obtained by comparison with the BFKL calculation.
This will be a subject of our future studies.

\section{Conclusions}

In the present letter we have discussed how the double-parton scattering
effects may contribute to large-rapidity-distance dijet correlations.
The present exploratory calculation has been performed in leading-order
approximation to understand and explore the general situation. This means that also
each step of DPS was calculated in collinear pQCD leading-order. We have shown 
that already leading-order calculation provides quite adequate
description of inclusive jet production when confronted with
recent results obtained by the ATLAS and CMS collaborations.
We have identified the dominant partonic pQCD subprocesses relevant for 
the production of jets with large rapidity distance.

We have concentrated ourselves on distributions in rapidity distance between
the most-distant jets in rapidity. The results of the dijet SPS
mechanism have been compared to the DPS mechanism. We have performed
calculations relevant for planned CMS analysis. The contribution of 
the DPS mechanism increases with increasing distance in rapidity between jets.
This is analogous to similar observations made already for the production 
of $c \bar c c \bar c$ \cite{Luszczak:2011zp,Maciula:2013kd,Hameren2014}
and $J/\psi J/\psi$ mesons \cite{Kulesza-Stirling,Baranov:2012re}. 
For comparison we have also shown some recent predictions of the Mueller-Navelet jets
in the LL and NLL BFKL framework from the literature.
For the CMS configuration our DPS contribution is smaller than 
the SPS dijet contribution even at high rapidity distances and
only slightly smaller than that for the NLL BFKL calculation known
from the literature.
The DPS final state topology is clearly different than that for the
SPS dijets (four versus two jets) which may help to disentangle the 
two mechanisms experimentally. Of course SPS three- and four-jet 
final states should be included in more refined analyses of
distributions in rapidity distance.

We have shown that the relative effect of DPS could be increased
by lowering the transverse momenta of jets but such measurements
can be difficult if not impossible. Alternatively one could study 
correlations of semihard pions distant in rapidity. Correlations 
of two neutral pions could be done, at least in principle, with 
the help of zero-degree calorimeters present at each main detectors 
at the LHC. This type of studies requires further analyses taking 
into account also hadronization effects.

The DPS effects are interesting not only in the context how they 
contribute to distribution in rapidity distance but \textit{per se}.
One could make use of correlations in jet transverse momenta,
jet imbalance and azimuthal correlations to enhance or lower the contribution
of DPS. Further detailed Monte Carlo studies are required to settle 
real experimental program of such studies.

\vspace{1cm}

{\bf Acknowledgments}

This work was supported in part by the Polish grant
DEC-2011/01/B/ST2/04535 as well as by the Centre for
Innovation and Transfer of Natural Sciences and Engineering Knowledge in
Rzesz{\'o}w.


\end{document}